\documentclass[reprint,twocolumn,superscriptaddress,showkeys,
nofootinbib,notitlepage,amsmath,amssymb,floatfix]{revtex4-2}

\pdfoutput=1
\usepackage{latexsym,amsmath,amssymb,lmodern,float,url}
\usepackage{natbib}
\usepackage{color}
\usepackage{multirow}
\usepackage{bm}
\usepackage{array}
\usepackage[normalem]{ulem}

\usepackage{mathtools}

\def\de{\delta^{\vphantom{1}}}
\def\bde{{\bar\delta}}
\def\qq{{q\bar q^\prime}}
\def\QQ{{Q\bar Q}}

\def\ccss{{c\bar{c}s\bar{s}}}

\def\ccqq{{c\bar{c}q\bar{q}^\prime}}

\def\bt{{\bar\theta}}

\def\h3{{\displaystyle{\frac 3 2}}}

\newcommand{\bbar}{\overline}

\begin{document}
\title{Diabatic Representation of Exotic Hadrons in the Dynamical Diquark Model}
\author{Richard F. Lebed}
\email{Richard.Lebed@asu.edu}
\author{Steven R. Martinez}
\email{srmart16@asu.edu}
\affiliation{Department of Physics, Arizona State University, Tempe,
AZ 85287, USA}
\date{July, 2022}

\begin{abstract}
We apply the diabatic formalism, an extension of the adiabatic
approximation inherent to the Born-Oppenheimer (BO) approach of
atomic physics, to the problem of mixing between exotic multiquark
hadrons and their nearby di-hadron thresholds.  The unperturbed BO
eigenstates are obtained using the dynamical diquark model, while the
diabatic calculation introduces a mixing potential between these
states and the threshold states.  We solve the resulting coupled
Schr\"{o}dinger equations numerically for hidden-charm tetraquarks of
both open and closed strangeness to obtain physical mass eigenvalues,
and explore the di-hadron state content and spatial extent of the
eigenstates.  As an explicit example, $X(3872)$ emerges with a
dominant $D^0 \bar D^{*0}$ component, but also contains a
considerable diquark-antidiquark component that can contribute
significantly to its radiative decay widths, and this component also
generates a full multiplet of other diquark-based exotic hadrons to
be compared with experiment.
\end{abstract}

\maketitle

\section{Introduction}

With the approach of the two-decade mark for experimental evidence of
heavy-quark exotic hadrons~\cite{Choi:2003ue}, the field remains at a
remarkable point: Well over 50 candidates have been observed with a high
degree of statistical significance, and yet no single theoretical
paradigm satisfactorily accounts for all of them~\cite{Lebed:2016hpi,
Chen:2016qju,Hosaka:2016pey,Esposito:2016noz,Guo:2017jvc,Ali:2017jda,
Olsen:2017bmm,Karliner:2017qhf,Yuan:2018inv,Liu:2019zoy,
Brambilla:2019esw,Chen:2022asf}.  Even so, several of these states lie
exceptionally close to di-hadron thresholds, the most
extraordinary example being the first one discovered, $X(3872)$:
\begin{equation} \label{eq:Xbind}
m_{X(3872)} \, – \, m_{D^0} \, – \, m_{D^{*0}} = -0.04 \pm 0.09 \
{\rm MeV} \, ,
\end{equation}
using the average mass value for each hadron in Eq.~(\ref{eq:Xbind})
tabulated by the Particle Data Group (PDG)~\cite{Zyla:2020zbs}.

No other hadron lies in such close proximity to a decay threshold,
suggesting a unique importance for the $D^0 {\bar D}^{*0}$ component
(the inclusion of charge conjugates being understood throughout)
within the state.  At minimum, a state so close to threshold exhibits
a number of ``universality’’ properties that depend only upon the
large $D^0 {\bar D}^{*0}$ scattering length~\cite{Braaten:2003he}.
More typically, $X(3872)$ has frequently been considered as a
hadronic molecule analogous to the deuteron $d$, which is largely
(but not entirely) bound by $\pi$ exchanges between its nucleon
components.  However, the 2.23~MeV $d$ binding energy is many times
larger than Eq.~(\ref{eq:Xbind}), and corresponds to a typical
hadronic size comparable to its root-mean-square (rms) charge radius
$\sqrt{\! \left < r_d^2 \right>} \!
= 2.13$~fm~\cite{Tiesinga:2021myr}.  By the same token, $X(3872)$
would be expected to be many femtometers across, larger than most
nuclei, and its physical observables would be utterly dominated by
long-distance $D^0$-${\bar D}^{*0}$ interactions.

Nevertheless, $X(3872)$ exhibits certain properties suggesting the
significance of its short-distance wave-func\-tion components.  Its
known decays to conventional charmonium ($J/\psi$ and $\chi_{c1}$)
account for more than 10\% of $\Gamma_{X(3872)}$, and one of its
observed radiative decay modes, $\gamma \psi(2S)$, has a branching
fraction of at least a few percent~\cite{Zyla:2020zbs}.  Since
charmonium rms charge radii are predicted from quark-potential models
to be significantly less than 1~fm, the large expected separation of
the charm quarks in the $D^0$-${\bar D}^{*0}$ pair would naively
predict much smaller branching ratios for these processes.  But
$X(3872)$ also shares the $J^{PC}$ quantum numbers  $1^{++}$ of the
yet-unseen conventional charmonium state $\chi_{c1}(2P)$ that
quark-potential models predict ({\it e.g.},
Ref.~\cite{Barnes:2005pb}) to lie several tens of MeV above
$m_{X(3872)}$.  As a result, $X(3872)$ has long been suggested to
contain a substantial core of conventional
$\chi_{c1}(2P)$~\cite{Suzuki:2005ha}, in part as a mechanism to
explain its surprising decay patterns.

Nevertheless, conventional charmonium is not the only short-distance
component available to $X(3872)$.  The valence quark content
$c\bar c u\bar u$ of $D^0 {\bar D}^{*0}$ allows for an alternative
binding mechanism: that of a diquark-antidiquark
$(cu)_{\bar{\bm 3}}(\bar c \bar u)_{\bm 3}$ pair, each one bound
through  the attractive color channel
${\bm 3} \otimes {\bm 3} \! \to \! \bar{\bm 3}$.

A number of other exotic candidates lie quite close to (within a few
MeV of) di-hadron thresholds, notably, $Z_c(3900)$ [$D {\bar D}^*$],
$Z_c(4020)$ [$D^* {\bar D}^*$], $P_c(4312)$ [$\Sigma_c {\bar D}$],
$P_c(4450)$/$P_c(4457)$ [$\Sigma_c {\bar D}^*$],
$Z_b(10610)$ [$B {\bar B}^*$], $Z_b(10650)$ [$B^* {\bar B}^*$], and
others.  In fact, all of these $Z$ meson states lie slightly
{\em above\/} the corresponding thresholds, arguing against a
traditional bound-state molecular picture.  And yet, one cannot deny
the significance of the proximity of the thresholds in these cases,
suggesting a special importance of those particular hadron pairs for
the exotic state.  On the other hand, some exotic candidates
[{\it e.g.}, the $Y$ states or $Z_c(4430)$] lack an obvious nearby
threshold.  A complete theoretical framework accommodating all of the
heavy-quark exotic candidates must therefore recognize the physical
significance of such nearby thresholds on their formation, mass, and
decay modes.

One formalism that has predicted a specific spectrum of multiquark
heavy-quark exotic hadrons is the
{\em dynamical diquark model}~\cite{Brodsky:2014xia,Lebed:2017min},
in which the formation of a di\-quark-antidiquark ($\de$-$\bde$)
exotic meson requires the color-nonsinglet diquark quasiparticles
$\de, \bde$ not to dissociate instantly into a di-meson pair, but
rather to persist as components of a single multiquark state
connected by a color flux tube.  Since each diquark is a color
triplet containing a heavy quark, the same potentials that are
computed on the lattice to describe heavy quarkonium and its hybrids
can be imported into Schr\"{o}dinger equations, which are solved
numerically to obtain the spectrum of $\de \bde$ eigenstates.  This
procedure has been performed both for the multiplet average
masses~\cite{Giron:2019bcs}, and for the detailed spectrum once
spin and isospin fine-structure effects are
included~\cite{Giron:2019cfc,Giron:2020fvd,Giron:2020qpb,
Giron:2021sla}.  Pentaquarks are handled
similarly~\cite{Giron:2021fnl}, by replacing the antidiquark $\bde$
with a color-triplet triquark $\bt \! \equiv \!
[\bar Q_{\bar {\bf 3}}
(q_1 q_2)_{\bar {\bf 3}}]_{\bf 3}$~\cite{Lebed:2015tna}.

This treatment of obtaining eigenstates of heavy (quasi-)particles
at separation $r$ connected by a static potential manifests the
well-known Born-Oppenheimer (BO) approximation~\cite{Born:1927boa}
from atomic physics.  Intrinsic to the approximation is the
assumption that the light degrees of freedom (d.o.f.) of the state
adjust instantly to changes in the configuration $\Gamma$ of the
heavy sources (an {\it adiabatic\/} approximation), and that the
eigenstates of the potential $V_\Gamma(r)$ thus derived from each
such configuration $\Gamma$ change gradually with $V_\Gamma(r)$.
However, when the value of $V_\Gamma (r)$ crosses the energy of a
di-hadron threshold, the physical eigenstates undergo a rapid level
crossing between a predominantly $\de \bde$ state and a predominantly
di-hadron state.  The mixing of configurations induces
{\it diabatic\/} changes to the system, and solving for its new
eigenstates requires a generalization beyond the strict BO limiting
case.

Such a diabatic formalism has been extensively developed in atomic
physics, and in recent years it has been standardized into textbook
form~\cite{Baer:2006}.  This approach provides a specific,
nonperturbative method for incorporating the mixing of
coupled-channel contributions from di-hadron states into the
calculations, and we briefly review the relevant formalism below.
The diabatic formalism was first applied to heavy-quark systems quite
recently in Ref.~\cite{Bruschini:2020voj}; in that case, the analysis
examined mixing of di-hadron thresholds with conventional
quarkonium, rather than with the 4-quark $\de \bde$ states of the
dynamical diquark model that are used here. 

The initial study of Ref.~\cite{Bruschini:2020voj} specifically
considers states lying below or just above a di-hadron threshold.  In
this work we adopt the same restriction, specifically to study the
lightest hidden-charm exotics of valence quark content
$c\bar c q\bar q^\prime$ (where $q^{(\prime)}$ are $u$ or $d$
quarks), $c\bar c s\bar s$, and $c\bar c q\bar s$.  ``Lightest'' in
this sense means the members of the (positive-parity) ground-state BO
multiplet $\Sigma^+_g(1S)$; observed candidates with (presumed)
corresponding flavor contents include $X(3872)$, $Y(4140)$, and
$Z_{cs}(4000)$, respectively.  Of course, the same methods can be
applied as well to orbitally excited multiplets such as
$\Sigma^+_g(1P)$ [containing, {\it e.g.}, $Y(4220)$], hidden-bottom
[{\it e.g.}, $Z_b(10610)$] states, and fully charmed
$c\bar c c\bar c$ [{\it e.g.}, $X(6900)$] states, and can also be
generalized to resonant states in order to study mass shifts and
strong-decay widths~\cite{Bruschini:2021cty}.

Here we focus on identifying the $\de$-$\bde$ and meson-meson content
(including distinct contributing partial waves) of the lightest
hidden-charm states, and extract interesting features such as the
expectation values $\left< r \right>$ and $\langle r^2 \rangle^{1/2}$
of the $\de$-$\bde$ separation $r$ in each mass eigenstate.  We show,
for example, that the wave function of $X(3872)$ is indeed dominated
by $D^0 {\bar D}^{*0}$, but not overwhelmingly so, and find that the
spectrum of states described by the original uncoupled dynamical
diquark model is not greatly disrupted by the existence of di-hadron
thresholds.

This paper is organized as follows.  In Sec.~\ref{sec:States} we
review the state notation for $\de \bde$ systems, focusing for now
only on the $\Sigma^+_g(1S)$ multiplet.  Section~\ref{sec:Form}
reviews the diabatic mixing formalism needed for the current set of
calculations.  Our numerical results appear in Sec.~\ref{sec:Res}, and in
in Sec.~\ref{sec:Concl} we summarize and indicate the next directions
for future calculations.

\section{States of the dynamical diquark  Model}
\label{sec:States}

The complete spectra of $\ccqq,c\bar c s\bar s$, $c\bar c q\bar s$,
and $c\bar c c\bar c$ states as $\de \bde$ eigenstates of the
dynamical diquark model are presented in Refs.~\cite{Lebed:2017min},
\cite{Giron:2020qpb}, \cite{Giron:2021sla}, and \cite{Giron:2020wpx},
respectively. In this paper, all relevant states are accommodated by
the lowest ($\Sigma^+_g$) Born-Oppenheimer (BO) potential, which
consists of the (light) gluon field in its ground state connecting
the heavy diquark [$\de \! \equiv (Qq)_{ \bar{\bf 3} }$]-antidiquark
[$\bde \! \equiv \! (\bar Q \bar q^\prime)_{\mathbf 3} $] or
diquark-triquark [$\bt \! \equiv \! (\bar Q_{\bar{\bf 3} }
(q_1 q_2)_{\bar{\bf 3}})]_{\bf 3}$ quasiparticles.  In all cases,
$\de, \bde, \bt$ are assumed to transform as color triplets (or
antitriplets), and each quasiparticle contains no internal orbital
angular momentum.

In this work, we consider $\QQ q\bar q^\prime$ states, in which
$q, \bar q^\prime$ may assume any of the flavors $\{ u,d,s \}$. The
classification scheme, regardless of the combination, begins with
6~states, here grouped by $J^{PC}$ quantum numbers.  This spectrum,
\begin{eqnarray}
J^{PC} = 0^{++}: & \ & X_0 \equiv \left| 0_\de , 0_\bde \right>_0 \,
, \ \ X_0^\prime \equiv \left| 1_\de , 1_\bde \right>_0 \, ,
\nonumber \\
J^{PC} = 1^{++}: & \ & X_1 \equiv \frac{1}{\sqrt 2} \left( \left|
1_\de , 0_\bde \right>_1 \! + \left| 0_\de , 1_\bde \right>_1 \right)
\, ,
\nonumber \\
J^{PC} = 1^{+-}: & \ & \, Z \  \equiv \frac{1}{\sqrt 2} \left( \left|
1_\de , 0_\bde \right>_1 \! - \left| 0_\de , 1_\bde \right>_1 \right)
\, ,
\nonumber \\
& \ & \, Z^\prime \equiv \left| 1_\de , 1_\bde \right>_1 \, ,
\nonumber \\
J^{PC} = 2^{++}: & \ & X_2 \equiv \left| 1_\de , 1_\bde \right>_2 \,
,
\label{eq:Swavediquark}
\end{eqnarray}
which specifies the full multiplet of $\Sigma^+_g$ $S$-wave states,
is written with the total $\de(\bde)$ spin denoted by
$s^{\vphantom\dagger}_{\de}(s_{\bde})$, and with the overall state
total spin signified by an outer subscript.  When needed, $9j$
angular momentum recoupling coefficients may be used to transform
these states to another spin basis.  For example, the transformation
coefficients to the basis of good total heavy-quark ($\QQ$) and
light-quark ($\qq$) spin read
\begin{eqnarray}
&&\left< (s_q \, s_{\bar q}) s_\qq , (s_Q \, s_{\bar Q}) s_\QQ
, S \, \right| \left. (s_q \, s_Q) s_\de , (s_{\bar q} \, s_{\bar Q})
s_\bde , S \right> \nonumber\\
&&=\left( [s_\qq] [s_\QQ] [s_\de] [s_\bde] \right)^{1/2}
\begin{Bmatrix}
s_q & s_{\bar q} & s_\qq \\
s_Q & s_{\bar Q} & s_\QQ \\ 
s_\de & s_\bde & S
\end{Bmatrix}
\, , \ \ \label{eq:9jTetra}
\end{eqnarray}
with $[s] \! \equiv \! 2s + 1$ denoting the multiplicity of a
spin-$s$ state.  Using Eqs.~(\ref{eq:Swavediquark}) and
(\ref{eq:9jTetra}), one may write
\begin{eqnarray}
J^{PC} = 0^{++}: & \ & X_0 = \frac{1}{2} \left| 0_\qq , 0_\QQ
\right>_0 + \frac{\sqrt{3}}{2} \left| 1_\qq , 1_\QQ \right>_0 \, ,
\nonumber \\
& & X_0^\prime = \frac{\sqrt{3}}{2} \left| 0_\qq , 0_\QQ
\right>_0 - \frac{1}{2} \left| 1_\qq , 1_\QQ \right>_0 \, , 
\nonumber \\
J^{PC} = 1^{++}: & \ & X_1 = \left| 1_\qq , 1_\QQ \right>_1 \, ,
\nonumber \\
J^{PC} = 1^{+-}: & \ & \, Z \; = \frac{1}{\sqrt 2} \left( \left| 
1_\qq , 0_\QQ \right>_1 \! - \left| 0_\qq , 1_\QQ \right>_1 \right)
\, , \nonumber \\
& \ & \, Z^\prime = \frac{1}{\sqrt 2} \left( \left| 1_\qq ,
0_\QQ \right>_1 \! + \left| 0_\qq , 1_\QQ \right>_1 \right) \, ,
\nonumber \\
J^{PC} = 2^{++}: & \ & X_2 = \left| 1_\qq , 1_\QQ \right>_2 \, .
\label{eq:SwaveQQ}
\end{eqnarray}

Further specifying the chosen combination of $\{ u,d,s \}$
light-quark flavors enlarges this set.  For example,  considering
combinations of $\{u,d\}$ alone expands the set to 12 states: 6 each
with $I \! = \! 0$ and $I \! = \! 1$, but which nonetheless maintain
spin structures in the forms of Eqs.~(\ref{eq:Swavediquark}) or
(\ref{eq:SwaveQQ}).  For the purposes of this work, we identify
states solely based upon total $J^{PC}$, effectively ignoring fine
structure due to isospin.  Additionally, we separately examine unique
flavor combinations of light quarks: $c\bar c s\bar s$,
$c\bar c q\bar s$, and $\ccqq$, where henceforth
$q,q^\prime \in \{u,d\}$.  The only sources of SU(3)$_{\rm{flavor}}$
dependence in these calculations arise through distinct explicit
diquark and meson masses.  

All states considered within this work are accommodated within the
ground-state BO multiplet $\Sigma^+_g(1S)$, but it is worth noting
that Ref.~\cite{Lebed:2017min} provides a classification of states in
higher multiplets such as $\Sigma^+_g(nP)$, as well as those with
excited-glue BO potentials such as $\Pi_u^+$.

\section{Diabatic Mixing Formalism}
\label{sec:Form}

In this work, we begin with the same construction as in the original
dynamical diquark model~\cite{Giron:2019bcs}.  That is, one separates
the light d.o.f.\ from the heavy d.o.f.\ by writing the Hamiltonian
as
\begin{equation} 
\label{eq:SepHam}
H=K_{\rm heavy} + H_{\rm light} =
\frac{\mathbf{p}^2}{2 \mu_{\rm heavy}} + H_{\rm light},
\end{equation}
such that the Schr\"{o}dinger equation now reads 
\begin{equation} \label{eq:SE}
\left( \frac{\mathbf{p}^2}{2 \mu_{\rm heavy}}+ H_{\rm light} - E \right)|\psi
\rangle = 0.
\end{equation}
Under the current analysis, ``light field'' refers to either just the
glue fields (in the case of a $\de$-$\bde$ state) or both glue and
exchanged light-quark fields (in the case of the meson-meson states).

We now implement the Ansatz that the states defined in
Sec.~\ref{sec:States} may appreciably mix with nearby meson-meson
thresholds sharing the same $J^{PC}$ quantum numbers, but assume that
the two types of states are clearly distinguishable away from the
thresholds.  Thus, one must determine and solve the multi-channel
Schr\"{o}dinger equation connecting the $\de \bde$ states to such
threshold states.  We closely follow the work of
Ref.~\cite{Bruschini:2020voj}, which carries out this process using
conventional quarkonium rather than $\de \bde$ states.  Applying the
diabatic expansion to the eigenstates of said Schr\"{o}dinger
equation yields~\cite{Bruschini:2020voj}
\begin{equation} 
\label{eq:AdExp}
|\psi \rangle = \sum_{i} \int d\mathbf{r}' \tilde \psi_i
(\mathbf{r}',\mathbf{r}_0) \: |\mathbf{r}' \rangle \:
|\xi_i(\mathbf{r}_0) \rangle,
\end{equation}
where $\mathbf{r}'$ denotes the separation of the heavy sources,
$\mathbf{r}_0$ is a freely set fiducial parameter, and
$|\xi_i \rangle$ are eigenstates of the light-field Hamiltonian.
Inserting Eq.~(\ref{eq:AdExp}) into Eq.~(\ref{eq:SE}) and applying
$\langle \xi_j (\mathbf{r}_0)| $ on the left-hand side produces
\begin{equation} 
\sum_{i} \left[ - \frac{\hbar^2}{2 \mu_{i}} \de_{ij}  \nabla ^2 +
V_{ji}(\mathbf{r,r_0})-E \de_{ji} \right] \tilde \psi_i (\mathbf{r,r_0}) = 0,
\end{equation}
with the {\em diabatic potential matrix\/} defined as
\begin{equation}
V_{ji}(\mathbf{r,r_0}) \equiv \langle \xi_j (\mathbf{r}_0)|
H_{\rm light} |\xi_i(\mathbf{r}_0) \rangle.
\end{equation}
We identify the $i \! = \! 0$ term with $\de \bde$ states, and
$i \! > \! 0$ terms with meson-meson states.

This result may be written more compactly in matrix notation as
\begin{equation} \label{eq:MatrixEqn}
[ \text K + \text V(\mathbf r)] \mathbf \Psi (\mathbf r) = \emph{E}
\, \mathbf \Psi (\mathbf r) \, ,
\end{equation}
with the parameter $\mathbf r_0$ implicit.  Neglecting interactions
between distinct meson-meson components, as is done in analogous
lattice-QCD studies~\cite{Bulava:2019iut}, the potential matrix then
becomes
\begin{equation}
\text V=
\begin{pmatrix}
V_{\de \bde}(\mathbf{r}) & V_{\rm mix}^{(1)}(\mathbf{r})  & \cdots &
V_{\rm mix \vphantom{\bbar M_2}}^{(N)}(\mathbf{r}) \\
V_{\rm mix}^{(1)}(\mathbf{r}) & 
V_{M_1 \bbar M_2}^{(1)}(\mathbf r) &
&
\\
\vdots
& & \ddots \\
V_{\rm mix \vphantom{\bbar M_2}}^{(N)}(\mathbf{r}) & & &
V_{M_1 \bbar M_2}^{(N)}(\mathbf r) \\
\end{pmatrix},
\end{equation}
with the kinetic-energy operator expressed as
\begin{equation}
\text K=
\begin{pmatrix}
-\frac{\hbar^2}{2\mu_{\de \bde}} & & & \\
& -\frac{\hbar^2}{2\mu^{(1)}_{M_1 \bbar M_2}} & & \\
& & \ddots & \\
& & & -\frac{\hbar^2}{2\mu^{(N)}_{M_1 \bbar M_2}} \\
\end{pmatrix}
\nabla^2,
\end{equation}
where omitted elements are zeroes.  Inspection of the $i>0$ diagonal
elements leads to the identification of
$V_{M_1 \bbar M_2}^{(i)}(\mathbf r)$ as simply being the energy
associated with that of the pure free $i^{\rm th}$ meson-meson state.
That is,
\begin{equation}
V_{M_1 \bbar M_2}^{(i)}(\mathbf r) = T_{M_1 \bbar M_2} \, ,
\end{equation}
where 
\begin{equation} \label{eq:Thresh}
T_{M_1 \bbar M_2} \equiv m^{\vphantom\dagger}_{M_1} \! +
m_{\bbar M_2} \, .
\end{equation}
For $S$-wave $\de \bde$ states, we identify
$V_{\de \bde}(\mathbf{r})$ as the uncoupled $\Sigma^{+}_g$
potential~\cite{Lebed:2017min}, with parameters that are calculated
on the lattice~\cite{Berwein:2015vca}.  Thus, we parametrize
\begin{equation}
\label{eq:sgmapot}
V_{\de \bde}(r)=- \frac{\alpha}{r} + \sigma r + V_0 + m_{\de} +
m_{\bde} \, ,
\end{equation}
where $\alpha,\sigma,$ and $V_0$ are $0.053 ~\text{GeV$\cdot$fm},
1.097 ~\text{GeV/fm},$ and $-0.180 ~\text{GeV}$, respectively.  For
each flavor sector, one also requires values for the $\de, \bde$
masses, and a corresponding list of di-hadron thresholds with
matching $J^{PC}$ quantum numbers must be identified.  One then needs
only to determine the appropriate form for the mixing potentials.
Reference~\cite{Bruschini:2020voj} argues for a Gaussian form, which
we also adopt here.  Explicitly, 
\begin{equation} \label{eq:Mixpot}
|V_{\rm mix}^{(i)} (r)| = \frac{\Delta}{2}
\exp \! \left\{ -\frac 1 2 \frac{\left[
V^{\vphantom\dagger}_{\de \bde}(r) -
T_{M_1 \bbar M_2 }^{(i)} \right]^2}{(\rho \sigma)^2} \right\} ,
\end{equation}
where $\sigma$ is the same string-tension parameter as in the
Cornell-like potential of Eq.~(\ref{eq:sgmapot}).  $\Delta$ is a free
parameter with units of energy indicating the strength of the mixing,
and $\rho$ is the radial scale for the level
crossing~\cite{Bruschini:2020voj}.

It is a useful first exercise to consider this procedure for a single
threshold.  The diabatic-potential matrix then reads
\begin{equation}
\text{V}(r) = 
\begin{pmatrix}
V^{\vphantom\dagger}_{\de \bde}(r) & V^{\vphantom\dagger}_{\rm mix}
(r) \\
V^{\vphantom\dagger}_{\rm mix} (r) & T_{M_1 \bbar M_2}
\end{pmatrix},
\end{equation}
where we have replaced $\mathbf{r}$ with its magnitude, since all
relevant potentials at this stage depend solely upon the diquark
separation distance.  The corresponding single-threshold
kinetic-energy operator becomes
\begin{equation}
\text K = 
\begin{pmatrix}
-\frac{\hbar^2}{2\mu_{\de \bde}} & \\
 & -\frac{\hbar^2}{2\mu_{M_1 \bbar M_2}} 
\end{pmatrix}
\nabla^2.
\end{equation}
The diabatic-potential matrix eigenvalues [denoted $V_- (r)$ and
$V_+ (r)$] and corresponding eigenvectors [$|\xi_- (r) \rangle$ and
$|\xi_+ (r) \rangle$] may be directly related to the $\de \bde$ and
di-meson threshold light-field eigenstates via a generic
transformation matrix:
\begin{equation}
\text{R}(r) = 
\begin{pmatrix}
\ \ \cos \theta(r) & \sin \theta(r) \\
-\sin \theta(r) & \cos \theta(r)
\end{pmatrix},
\end{equation}
such that
\begin{equation}
\text{R}(r) \text{V}(r) \text{R}^\dagger (r) = \text{diag}
\left\{ V_-(r), V_+(r) \right\}  \, .
\end{equation}
Here, $\theta$ is identified as the mixing angle between the
$\de \bde$ and meson-meson terms.  Since $\text{R}(r)$ diagonalizes
$\text{V}(r)$, one can deduce an expression for $\theta$ in terms of
its matrix elements~\cite{Bruschini:2020voj}:
\begin{equation} \label{eq:mixingangle}
\theta (r) = \frac{1}{2} \arctan \left(
\frac{2V_{\rm mix}(r)}{T_{M_1 \bbar M_2} -
V^{\vphantom\dagger}_{\de \bde}(r)} \right) .
\end{equation}
As an example, we plot $\theta$ for $\de (\bde) = cq \,
(\bbar{cq}')$ and $M_1 \bbar M_2 = D \bar D^*$ in
Fig.~\ref{fig:mixingangle} using phenomenologically viable values of
$\Delta$ and $\rho$ [see Eq.~(\ref{eq:rhodelta})], as obtained below.

\begin{table*}[ht]
\caption{Calculated eigenvalues and component-state admixtures for
the $\ccqq$ sector obtained from solving Eq.~(\ref{eq:MatrixEqn}) for
specific $J^{PC}$ numbers. Also presented are the expectation values
$\langle r \rangle$ of the radial coordinate $r$ (corresponding to
$\delta$-$\bde$ separation) as well as $\langle r^2 \rangle^{1/2}$
for each state.  Suppressed entries indicate contributions that are
individually $< \! 1\%$.}
\setlength{\tabcolsep}{9pt}
\renewcommand{\arraystretch}{1.2}
\begin{tabular}{c  c  c c c c c  c  c} 
 \hline\hline
 $J^{PC}$ & $E \ ({\rm MeV})$ & $\de \bde$ & $D \bar D^*$ &
 $D_s \bar D_s$ & $D^* \bar D^*$  & $D_s^* \bar D_s^*$ &
 $\langle r \rangle ({\rm fm})$ & $\langle r^2 \rangle^{1/2}
 ({\rm fm})$ \\
 \hline
 $0^{++}$ & 3905.4  &  63.0\% & &  27.4\%  &
  8.4\% & 1.2\% & 0.596 & 0.605 \\ 
 $1^{++}$ & 3871.5  &  8.6\% &  91.4\% &  & & &
 4.974 & 5.459  \\
 $2^{++}$ & 3922.3  &  83.1\% & &  1.5\% &
  13.9\% & 1.5\% & 0.443 & 0.497 \\
 \hline\hline
\end{tabular}
\label{tab:ccqq}
\end{table*}

\section{Results}
\label{sec:Res}
\subsection{$\ccqq$ Sector}

\begin{figure}[t]
    \centering
    \includegraphics[scale=0.36]{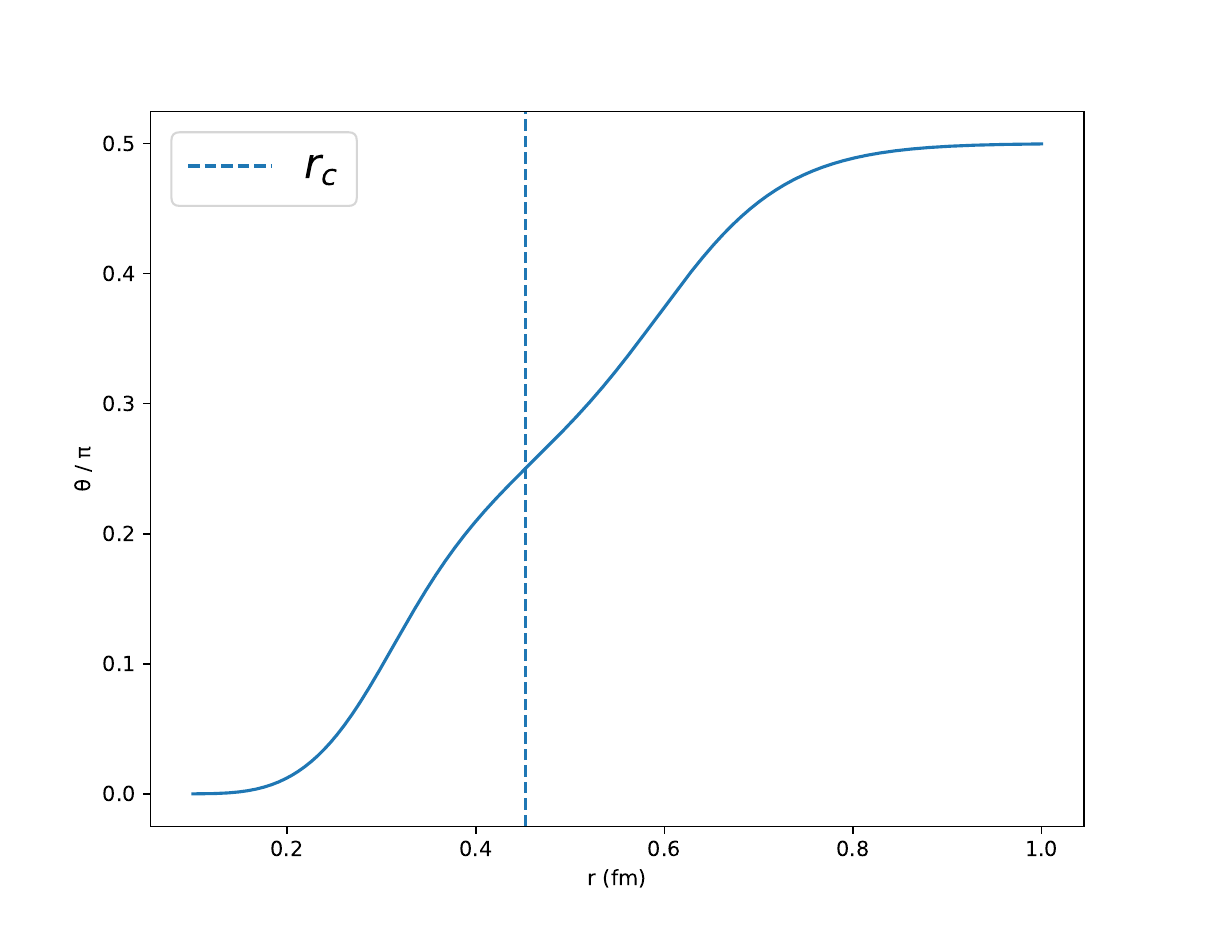}
    \caption{The mixing angle $\theta(r)$
    [Eq.~(\ref{eq:mixingangle}), using Eqs.~(\ref{eq:Mixpot}) and
    (\ref{eq:rhodelta})] between $\de \bde = c\bar c q q'$  and
   $D \bar D^*$ in the $1^{++}$ channel, as a function of
    heavy-source radial separation $r$.  The dashed line shows the
    critical radius $r_c$, {\it i.e.}, the separation at which
    $V_{\delta \bde}(r) = T_{D \bar D^*}$, using
    Eqs.~(\ref{eq:Thresh})--(\ref{eq:sgmapot}) and (\ref{eq:mcq}).}
    \label{fig:mixingangle}
\end{figure}

In order to obtain values for the $\Delta$ and $\rho$ parameters in
the mixing potentials, we fit them such that the lowest $1^{++}$
eigenvalue in the $\ccqq$ sector is near the measured value
$m_{X(3872)} \! = \! 3871.7$~MeV\@.  Of course, a range of
$\Delta, \rho$ value pairs achieves this result.  Following the
approach of  Ref.~\cite{Bruschini:2020voj}, we select a pair that
produces the correct asymptotic behavior for the single-threshold
mixing angle $\theta(r)$, as modeled in Fig.~\ref{fig:mixingangle}.
Specifically, as supported by unquenched lattice-QCD
calculations~\cite{Bali:2005}, we require that
$\theta \! \to \! \pi/2$ rapidly for $r \! > \!
r_c$, the value at which the $\de \bde$ Cornell-like potential
Eq.~(\ref{eq:sgmapot}) crosses the $M_1 \bbar M_2$ threshold
Eq.~(\ref{eq:Thresh}).  According to Eq.~(\ref{eq:mixingangle}), the
$\Delta, \rho$ pair must also produce the nonzero value
$\theta \! = \! \pi/4$ at $r \! = \! r_c$, indicating a nontrivial
mixing of the $\de \bde$ and meson-meson components.  The
specific values used here are
\begin{equation} \label{eq:rhodelta}
\Delta_{\de \bde}=0.300 \: \text{GeV} \, , \;
\rho_{\de \bde}=0.185 \: \text{fm} \, .
\end{equation}
Again, these values are not unique, but rather serve as proof of
principle for the approach.  The diquark mass is then the only
remaining free parameter, with its value taken as reported in
Ref.~\cite{Giron:2021sla}:
\begin{equation} \label{eq:mcq}
m_{\de = (cq)}=m_{\bde = (\bar c \bar q)}=1.9271 \; \text{GeV}.
\end{equation}

One may then numerically solve Eq.~(\ref{eq:MatrixEqn}) as a coupled
set of equations.  We follow the procedure described in Sec.~IV.F of
Ref.~\cite{Bruschini:2020voj}, solving for the lowest bound states.
The resulting eigenvalues and their corresponding state mixtures are
collected in Table~\ref{tab:ccqq}.  Note the omission of a
$J^{PC}=1^{+-}$ entry in Table~\ref{tab:ccqq} as compared with
Eqs.~(\ref{eq:Swavediquark}) or (\ref{eq:SwaveQQ}); while the
diabatic formalism allows for the formation of $C \! = \! -1$
eigenstates, the di-hadron thresholds alone provide no mechanism to
lift the degeneracy with the $1^{++}$ eigenstate.\footnote{In the
full dynamical diquark model, this degeneracy is lifted by operators
that couple to individual diquark spins~\cite{Giron:2019cfc,
Giron:2020qpb}.}  Notably, we find that the $1^{++}$ eigenstate,
which we associate with $X(3872)$, consists primarily of $D \bar D^*$
content (an idea known for quite some time, {\it e.g.},
Refs.~\cite{Suzuki:2005ha,Hosaka:2016pey,Yamaguchi:2019vea}),
consistent with expectations given its proximity to that threshold.
In contrast, we find that the other eigenstates ($0^{++}$ and
$2^{++}$) are primarily of $\de \bde$ content.

Using the results of this formalism, it is also possible to calculate
certain transition rates, which of course provide numerous
predictions for comparison with experimental results.  Broadly
accepted techniques for calculating the decays of $\de\bde$ to
$Q\bar Q$ states do not yet exist in the literature, but until such
methods are robustly developed, we can at least perform exploratory
studies using analogues of known expressions.  Here, we focus
specifically on the radiative transition of the $J^{PC}=1^{++}$
eigenstate to $J/\psi$ and $\psi(2S)$.  For E1 and M1 transitions of
states within the dynamical diquark model, substantial work has
already been performed in Ref.~\cite{Gens:2021wyf}.   There, a
standard equation for E1 partial widths for the process
$i \! \to \! \gamma f$ is adapted to the case of exotic-to-exotic
transitions, and it may be recast for the present case as
\begin{widetext}
\begin{equation}\label{eq:trans}
\Gamma_{\rm E1} \left( n^{2s_{Q \bar Q} + 1}(J_{q \bar q})_{J}
\rightarrow n^{\prime  {2s+1}}(L^{\prime})_{J^{\prime}} +
\gamma \right)=\frac{4}{3} C_{fi} \,
\delta_{s^{\vphantom\prime}_{Q \bar Q}s^{\prime}_{Q \bar Q}}
Q^{2}_{\delta} \alpha \, |
\langle \psi_f | r | \psi_i \rangle |^2 E^{3}_{\gamma}
\frac{E_{f}^{Q \bar Q}}{M_{i}^{Q\bar Q q\bar q}} \, ,
\end{equation}
\end{widetext}
where 
\begin{equation}
C_{fi} \equiv {\rm max} \left( J_{q \bar q} , L^{\prime} \right)
 (2J^{\prime}+1) 
\begin{Bmatrix}
L^{\prime} & J^{\prime} & s_{Q \bar Q} \\
J & J_{q \bar q} & 1
\end{Bmatrix}^2 \, .
\end{equation}
In this expression, $\alpha$ is the QED fine-structure constant,
$E_{\gamma}$ is the photon energy (measured in the rest frame of the
decaying exotic initial state $Q\bar Q q \bar q$ of mass
$M_i^{Q\bar Q q \bar q}$, in which the conventional quarkonium final
state $Q\bar Q$ recoils with energy $E_f^{Q\bar Q}$), and
$Q_{\de} \! = \! Q_Q \! + Q_q$ is the diquark charge in units of
proton charge (meaning that the diquark is treated as a single
quasiparticle coupling to the photon).  Furthermore,
$s^{(\prime)}_{Q\bar Q}$ denotes the total heavy-quark spin
eigenvalue for the initial (final) state, $n^{(\prime)}$ indicates
the principal quantum number for the initial (final) state,
$J_{q\bar q}$ ($L^\prime$) represents the total angular momentum
carried by the light d.o.f.\ in the initial (final) state, and $J$
($J^\prime$) is the total angular momentum eigenvalue of the initial
(final) state, respectively.

The original form of Eq.~(\ref{eq:trans}) refers solely to E1
transitions between two states with the same two fundamental charged
constituents, which together generate an electric dipole moment.
This expression is reasonable (and indeed is commonly used) for
conventional quarkonium, albeit often including relativistic
corrections that are outlined in Ref.~\cite{Gens:2021wyf}.  However,
several precarious assumptions enter into the use of
Eq.~(\ref{eq:trans}) for the transition of a $\de \bde$ to a
$Q\bar Q$ state.  The first is the use of a diquark as a single
charged quasiparticle to which a photon couples; this premise was
explored in Ref.~\cite{Gens:2021wyf} along with alternative
hypotheses, such as the incoherent coupling of the photon to the
component quarks of the diquark, leading to different $O(1)$ factors
taking the place of $Q_\delta^2$ in the analogue to
Eq.~(\ref{eq:trans}).  More significantly, however,
Ref.~\cite{Gens:2021wyf} applied Eq.~(\ref{eq:trans}) only to the
case of transitions between two $\de \bde$ states, again treating the
diquarks as single compact quasiparticles.  Applying it to
$\de \bde \! \to Q\bar Q$ transitions is much more questionable
since, among other possible objections, such a transition requires
the annihilation of the $q \bar q$ pair.  At minimum,
Eq.~(\ref{eq:trans}) must at least be modified to accommodate the
Okubo-Zweig-Iizuka (OZI) suppressed amplitude $q\bar q \! \to \! g$
in this process.  While a proper treatment of the issue lies outside
the scope of this work, we may at least note that the magnitude
$\epsilon$ of the OZI-suppression is expected to be significantly
$< \! 1$, and so $\Gamma_{\rm E1}$ of Eq.~(\ref{eq:trans}) is
diminished by a factor $\epsilon^2$.  One may expect $\epsilon$ to
depend upon both the radial excitation number $n$ of the initial
state, as well as the $q\bar q$ spin state.

The most sensitive term to calculate in Eq.~(\ref{eq:trans}) is the
overlap $| \langle \psi_f | r | \psi_i \rangle |^2$ of the initial
and final radial wave functions weighted by $r$, the characteristic
spatial separation of the heavy sources.  Note especially that
$\psi_i$ ($\psi_f$) is a $\de \bde$ ($Q\bar Q$) state, so that
computing this amplitude properly certainly requires more than
the simple evaluation of wave-function overlap performed here.
Recall that the full exotic eigenstate in this analysis is taken to
be a mixture of both $\de \bde$ and $M_1 \overline{M}_2$ components.
However, the expected relative separation of the mesons in the
$D^0 \bar D^{*0}$ component of $X(3872)$ is expected to be much
greater than that of a pure $\de \bde$ state.  Explicitly, using the
results of Table~\ref{tab:ccqq}, we find that the $D^0 \bar D^{*0}$
component alone should have $\langle r \rangle \! = \! 5.092$~fm, in
agreement with the crude estimate provided by the Bohr-radius
analogue
$r \! \sim \! \frac{1}{\sqrt{2 \mu E_{\rm bind.}}}$.\footnote{The
specific $O(1)$ coefficient in this expression arises from using the
literal Bohr radius for $r$.  A different $O(1)$ coefficient arises
when computing $\langle r^m \rangle^{1/m}$, but in any case one
obtains a result for $X(3872)$ of multiple fm.}  Conversely, the
value of $\langle r \rangle$ for the pure $\de (\bde) = cq \,
(\bar c \bar q^\prime)$ component is only 0.361~fm.  Since the decays
$X(3872) \! \to \! \gamma \psi$ require the annihilation of the
light $q\bar q$ pair in $X(3872)$, one naively expects that the
$\de \bde$ component should dominate in this process.  Let us test
this expectation.

Once the fractional $\de \bde$ content in the $1^{++}$ eigenstate
is known, one may calculate the E1 partial width from
Eq.~(\ref{eq:trans}).  Using the potential of Eq.~(\ref{eq:sgmapot})
and numerical inputs for its parameters given there, and adopting
$m_c \! = \! 1840$~MeV~\cite{Bruschini:2020voj} in place of
$m_{\de}$, we compute the radial wave functions of $J/\psi$,
$\psi(2S)$, and $\chi_{c1}(2P)$ treated as pure conventional
charmonium states.  We also solve for the $(1S)$ $\de \bde = cq
\, (\bar c \bar q )$ wave function, which was an essential ingredient
in Ref.~\cite{Gens:2021wyf}. Introducing these values into
Eq.~(\ref{eq:trans}), we find
\begin{eqnarray} \label{eq:etrans}
\Gamma_{\text E1}[ X(3872) \rightarrow \gamma J/\psi ]
& = & 469 \, \epsilon^2 \, \text{keV}, \nonumber \\
\Gamma_{\text E1}[ X(3872) \rightarrow \gamma \, \psi(2S) ]
& = & 1.56 \, \epsilon^2 \, \text{keV},
\end{eqnarray}
while
\begin{eqnarray} \label{eq:ctrans}
\Gamma_{\text E1}[ \chi_{c1}(2P) \rightarrow \gamma J/\psi ]
& = & {\rm 20.8} \, \text{keV}, \nonumber \\
\Gamma_{\text E1}[ \chi_{c1}(2P) \rightarrow \gamma \, \psi(2S) ]
& = & 78.0 \, \text{keV}.
\end{eqnarray}
We note that the results of Eqs.~(\ref{eq:ctrans}) are comparable to
the predictions of 71~keV and 95~keV, respectively,
from Ref.~\cite{Swanson:2004pp} [which also uses
Eq.~(\ref{eq:trans}), but with somewhat different numerical values
for the matrix elements].

On the other hand, Eq.~(\ref{eq:etrans}) predicts (setting
$\epsilon \! \to \! 1$) wildly different $X(3872)$ radiative widths
than the current PDG averages~\cite{Zyla:2020zbs},
$10.1 \! \pm \! 4.7$~keV for $J/\psi$ and $54 \pm 25$~keV for
$\psi(2S)$.  The PDG separately gives an average for their ratio:
\begin{equation} \label{eq:PDGR}
R \equiv \frac{\Gamma [ X(3872) \to \gamma \psi(2S) ]}
{ \Gamma [ X(3872) \to \gamma J/\psi ]} =  2.6 \pm 0.6  \, ,
\end{equation}
while the corresponding ratio from Eqs.~(\ref{eq:etrans}), taking the
two values of $\epsilon$ equal, is only $3.3 \times 10^{-3}$.
Interestingly, this number approximately equals an early estimate for
$R$ based upon a molecular-model calculation using vector-meson
dominance~\cite{Swanson:2004pp}.  However, a modern molecular-model
calculation employing an effective Lagrangian supports values of $R$
in a typical range of order several tenths~\cite{Guo:2014taa}, and
these predictions are numerically stable when a small admixture of
the conventional-charmonium state $\chi_{c1}(2P)$ is
included~\cite{Cincioglu:2019gzd}.

The origin of such a small predicted value for $R$ in the dynamical
diquark model can be traced to the fact that the $\de \bde$
component of $X(3872)$ is a ground-state ($n \! = \! 1$)
radial mode like $J/\psi$, while $\chi_{c1}(2P)$ is a first-excited
($n \! = \! 2$) radial mode like $\psi(2S)$.
Indeed, our calculated idealized $\de \bde$ wave function has almost
exactly the same size as our calculated $c\bar c$ $J/\psi$ wave
function, leading to almost complete overlap and a large transition
to $J/\psi$.  In a more realistic treatment including a finite
diquark $(cq)$ size~\cite{Giron:2019cfc}, one expects this overlap
enhancement to be muted.  Nevertheless, one still expects the
phase-space advantage of $J/\psi$ over $\psi(2S)$ to generate a
rather small $R$ value in this model.  If a value of $R$
substantially larger than 1 persists in the data, we conclude that
one must have some underlying preferential coupling to $\psi(2S)$
over $J/\psi$, either from a significant $\chi_{c1}(2P)$ component in
$X(3872)$, from an enhanced ratio of $\psi(2S)$ to $J/\psi$
effective-theory couplings as suggested in Ref.~\cite{Guo:2014taa},
or from an enhanced wave-function overlap of $\de \bde$ to the
spatially larger $\psi(2S)$ state.

The overall size of the prediction for $\Gamma [X(3872) \! \to \!
\gamma J/\psi]$ from the first of Eqs.~(\ref{eq:etrans}) also points
to the necessity of $\epsilon$ being no larger than about $1/6$,
which is not an unreasonable expectation for an OZI-suppressed
amplitude.

Lastly, in a model for $X(3872)$ with not just $\de \bde$ but also
$D^0 \bar D^{*0}$ components [and possibly $\chi_{c1}(2P)$ as well],
the full value for its radiative decay widths can include contributions from
more than one source.  In that case, it becomes crucial to determine
the relative phase of each contribution, since interference may be
critical to obtaining a physically accurate result.  This effect in a
mixed $D^0 \bar D^{*0}$-$\chi_{c1}(2P)$ model was considered in
Ref.~\cite{Cincioglu:2019gzd}.

\subsection{$\ccss$ and $c\bar c q\bar s$ Sectors}

In the $\ccss$ sector, one encounters an additional free parameter,
the $(cs)$ diquark mass.  We fix this mass so that the $0^{++}$
energy eigenvalue matches the $X(3915)$ mass~\cite{Zyla:2020zbs}.
This choice produces the result
\begin{equation}
m_{\de=(cs)}=m_{\bde=(\bar c \bar s)}=1.9450 \; \text{GeV}.
\end{equation}
Note that this value only slightly exceeds $m_{\de=(cq)}$ given in
Eq.~(\ref{eq:mcq}), despite containing a heavier $s$ quark; in
particular, it is substantially smaller than the value (2.080~GeV)
obtained from the analysis of Ref.~\cite{Giron:2021sla}, because that
work (unlike here) incorporates spin-splitting fine structure for the
$\Sigma^+_g(1S)$ $\de \bde$ multiplet.  The remaining eigenstates may
then be calculated, and are presented in Table~\ref{tab:ccss}.
Remarkably, the $1^{++}$ and $2^{++}$ states exhibit rather little
$D^{(*)}_{(s)} \bar D^{*}_{(s)}$ content, while the $0^{++}$ state
shows significant mixing with $D_s \bbar D_s$.

\setlength{\tabcolsep}{9pt}
\renewcommand{\arraystretch}{1.2}
\begin{table*}[ht!]
\caption{The same as in Table~\ref{tab:ccqq}, for the $\ccss$ sector.}
\begin{tabular}{c  c  c c c c c  c  c }
\hline\hline

$J^{PC}$ & $E \ ({\rm MeV})$ &  $\de \bde$ & $D_s \bar D_s$ &
$D^* \! \bar D^*$ & $D_s \bar D_s^*$ & $D_s^* \bar D_s^*$ &
$\langle r \rangle ({\rm fm})$ & $\langle r^2 \rangle^{1/2}
({\rm fm})$ \\
\hline
$0^{++}$ & 3922.0 &  50.6\% &  39.9\% &
 8.3\% & & 1.2\% & 0.627 & 0.757 \\
\hline
$1^{++}$ & 3978.7 &  86.9\% & &  1.5\% &
 11.0\% & & 0.421 & 0.470  \\
\hline
$2^{++}$ & 3998.7 &  95.2\% & &  &
 1.5\% & 3.3\% & 0.394 & 0.437  \\
\hline\hline
\end{tabular}
\label{tab:ccss}
\end{table*}
\begin{table*}[ht!]
\caption{The same as in Tables~\ref{tab:ccqq} \& \ref{tab:ccss}, for
the $c\bar c q\bar s$ sector.}
\begin{tabular}{ c  c  c c c c  c  c }
\hline\hline
$J^P$ & $E \ ({\rm MeV})$ & 
$\de \bde$ & $D^* \! \bar D_s$ & $D \bar D_s^*$ & $D^* \! \bar D_s^*$
& $\langle r \rangle ({\rm fm})$ & $\langle r^2 \rangle^{1/2}
({\rm fm})$ \\
\hline
$0^+$ & 3983.5 &  93.2\% & & &  6.8\% &
0.397 & 0.440 \\
\hline
$1^+$ & 3914.5 &  66.7\% &  16.5\%  &
 15.9\% && 0.482 & 0.548 \\
\hline
$2^+$ & 3962.1 &  90.4\% &  1.8\% &
 1.8\% &  6.0\% & 0.411 & 0.460 \\
\hline\hline
\end{tabular}
\label{tab:ccqs}
\end{table*}

The calculations for $c\bar c q \bar s$ states follow entirely from
the previously determined parameters, and are presented separately in
Table~\ref{tab:ccqs}.  This sector in particular contains a unique
threshold structure, in that the $D^* \! \bar D_s$ and $D \bar D^*_s$
thresholds differ by less than 4~MeV, for each fixed value of total
charge for the meson pair.  Our calculations show that thresholds
extremely close to each other tend to evenly share state content
(assuming that the allowed $\ell$ quantum numbers are the same).

In all three sectors, whenever $\de \bde \, (\ell \! = \! 0)$ content
dominates, we furthermore find a preferential coupling to $S$-wave
meson-meson combinations over $D$-wave combinations, even if the
latter threshold is closer to the mass eigenvalue.  This result is
exhibited by the $\ccqq ~ 2^{++}$, $\ccss ~1^{++}$, and
$c \bar c q \bar s ~2^+$ states.  The $\ccqq ~ 2^{++}$ state is the
most extreme of these, preferring to couple more strongly to
$D^*_s \bar D^*_s$ over $D_s \bar D_s$, despite the $\sim $280~MeV
difference.  This behavior is further exhibited by the results for
individual thresholds; for example, in the $2^{++} \ccqq$ state, the
distinct $D^* \bar D^*$ channels with $\ell=0$ and $\ell=2$ contribute
$11.6\%$ and $1.2\%$, respectively.  The expected suppression
associated with the near-orthogonal overlap of an $S$-wave $\de \bde$
state and a $D$-wave meson-meson state provides a natural
explanation.

We also note that the expectation values $\left< r \right>$ (or
$\langle r^2 \rangle^{1/2}$) for each state increase with increasing
meson-meson content, consistent with predictions of
$\left< r \right>$ for hadron molecular states~\cite{Liu:2008fh}.  As
discussed above, the $1^{++} \ccqq$ state exhibits a several-fm value
for $\left< r \right>$, as is expected for a pure $D \bar D^*$
molecular state with a binding energy of tenths of an
MeV~\cite{Liu:2008fh}.

\section{Conclusions}
\label{sec:Concl}

The introduction into a diquark-based model---in this paper, the
variant known as the dynamical diquark model---of effects caused by
the proximity of some of its eigenstates to di-hadron thresholds
creates a model for multiquark exotic hadrons that combines the best
features of both diquark and hadron-molecular models.  One universal
model can then incorporate exotic candidates that lie quite close to
such thresholds, but which still maintain a close physical connection
to other exotic candidates with no obvious di-hadron interpretation.

Coupled-channel calculations between bound states and thresholds have
of course been carried out many times in the past, originally in the
context of atomic and molecular physics, but more recently for
exotic-hadron candidates as well.  The so-called diabatic approach
used here earns distinction as the natural (and rigorous)
generalization of the Born-Oppenheimer approximation used to compute
the unperturbed spectrum of heavy-heavy systems such as quarkonium
and its hybrid excitations, and in this work for the first time, the
diquark-antidiquark states of the dynamical diquark model.

The precise functional form for the mixing potential between
unperturbed states and thresholds used here is phenomenological in
origin, but it is motivated by lattice-QCD results involving the
breaking of the color flux tube between the heavy sources.  Clearly,
advances in lattice simulations can be incorporated into improved
future calculations.

We find, using parameters from the most recent analysis in the
dynamical diquark model, that the famous $X(3872)$ can naturally
contain a dominant component of $D^0 \bar D^{*0}$ and yet originate
as the unique isosinglet $1^{++}$ member of the lightest multiplet of
hidden-charm diquark-antidiquark states.  Masses of all the remaining
members of the multiplet are then predicted, along with their
di-hadron content generated by nearby thresholds.  Since the original
model has been extended to study hidden-charm, hidden-strangeness and
open-strange states as well, we also present calculations for those
flavor sectors.  General results include the observation that
$S$-wave thresholds always produce larger effects than $D$-wave
thresholds, even if the latter are substantially closer to the
mass eigenvalue, and that coincident di-hadron thresholds with
different arrangements of the light flavors in the hadrons have
comparable effects.

We also use our results on state content to calculate the radiative
decay widths for $X(3872) \! \to \! \gamma J/\psi$ and
$\gamma \psi(2S)$, which provides crucial information on the
short-distance components of $X(3872)$, since its di-hadron component
is spatially much larger than that of its diquark-antidiquark
component.  Interestingly, we obtain a $\gamma \psi(2S)$ width much
smaller than the current measured value, due not just to the fact
that the compact component amounts to only about 10\% of the full
state, but also that the overlap of the $1S$ diquark-antidiquark and
$2S$ $c\bar c$ states is small.  We conclude that obtaining the full
radiative width may require including a comparable contribution from
the diffuse di-hadron component, or even from a $\chi_{c1}(2P)$
component (which can also be included in the diabatic formalism).

These initial calculations, while quite encouraging, remain quite
incomplete.  First, no spin- or flavor-dependent effects (besides
explicit differences in the diquark mass) have been incorporated into
the unperturbed side of this calculation.  The Hamiltonian for the
full dynamical diquark model contains spin-spin and spin-isospin
operators that, for example, distinguish masses of $X(3872)$,
$Z_c(3900)$, and $Z_c(4020)$.  In this calculation, the
only mass splittings arise from the explicit differences of
$D^{(*)}_{(s)}$ masses.  One major future research thrust is the
inclusion of explicit fine-structure effects into the initial
unperturbed states.

A second possible improvement involves the treatment of the
threshold contribution.  Di-hadron molecular models have been alluded
to several times in this paper, but the actual treatment of the
di-hadron state uncoupled from the diquark-antidiquark state,
according to Eq.~(\ref{eq:Thresh}), is simply that of two free
hadrons.  Should it be desirable to regard the pair as forming a true
di-hadron molecule, then directly replacing Eq.~(\ref{eq:Thresh})
with a mildly attractive {\it e.g.}, meson-exchange) potential wouldbe straightforward in this formalism.

Finally, the diabatic formalism as presented here strictly only
applies to states either below or not too far above significant
di-hadron thresholds.  Some of the exotic candidates [{\it e.g.},
$Z_c(4430)$] lie rather far from relevant thresholds, and
consequently have large decay widths.  Such broad resonances should
properly be treated as poles in di-hadron scattering amplitudes.
However, the diabatic formalism has been developed, in the case of
mixing with conventional quarkonium, to include the calculation of
strong decay widths~\cite{Bruschini:2021cty} and di-hadron scattering
amplitudes~\cite{Bruschini:2021ckr}.  These methods can be
immediately adapted to the case of mixing with diquark-antidiquark
states, and will also be incorporated into future work.

\vspace{1em}

\begin{acknowledgments}
This work was supported by the National Science Foundation (NSF) under 
Grants No.\ PHY-1803912 and PHY-2110278.
\end{acknowledgments}

\bibliographystyle{apsrev4-2}
\bibliography{Diabatic}

\begin{thebibliography}{42}%
\makeatletter
\providecommand \@ifxundefined [1]{%
 \@ifx{#1\undefined}
}%
\providecommand \@ifnum [1]{%
 \ifnum #1\expandafter \@firstoftwo
 \else \expandafter \@secondoftwo
 \fi
}%
\providecommand \@ifx [1]{%
 \ifx #1\expandafter \@firstoftwo
 \else \expandafter \@secondoftwo
 \fi
}%
\providecommand \natexlab [1]{#1}%
\providecommand \enquote  [1]{``#1''}%
\providecommand \bibnamefont  [1]{#1}%
\providecommand \bibfnamefont [1]{#1}%
\providecommand \citenamefont [1]{#1}%
\providecommand \href@noop [0]{\@secondoftwo}%
\providecommand \href [0]{\begingroup \@sanitize@url \@href}%
\providecommand \@href[1]{\@@startlink{#1}\@@href}%
\providecommand \@@href[1]{\endgroup#1\@@endlink}%
\providecommand \@sanitize@url [0]{\catcode `\\12\catcode `\$12\catcode
  `\&12\catcode `\#12\catcode `\^12\catcode `\_12\catcode `\%12\relax}%
\providecommand \@@startlink[1]{}%
\providecommand \@@endlink[0]{}%
\providecommand \url  [0]{\begingroup\@sanitize@url \@url }%
\providecommand \@url [1]{\endgroup\@href {#1}{\urlprefix }}%
\providecommand \urlprefix  [0]{URL }%
\providecommand \Eprint [0]{\href }%
\providecommand \doibase [0]{https://doi.org/}%
\providecommand \selectlanguage [0]{\@gobble}%
\providecommand \bibinfo  [0]{\@secondoftwo}%
\providecommand \bibfield  [0]{\@secondoftwo}%
\providecommand \translation [1]{[#1]}%
\providecommand \BibitemOpen [0]{}%
\providecommand \bibitemStop [0]{}%
\providecommand \bibitemNoStop [0]{.\EOS\space}%
\providecommand \EOS [0]{\spacefactor3000\relax}%
\providecommand \BibitemShut  [1]{\csname bibitem#1\endcsname}%
\let\auto@bib@innerbib\@empty
\bibitem [{\citenamefont {Choi}\ \emph {et~al.}(2003)\citenamefont {Choi} \emph
  {et~al.}}]{Choi:2003ue}%
  \BibitemOpen
  \bibfield  {author} {\bibinfo {author} {\bibfnamefont {S.}~\bibnamefont
  {Choi}} \emph {et~al.} (\bibinfo {collaboration} {Belle Collaboration}),\
  }\href {https://doi.org/10.1103/PhysRevLett.91.262001} {\bibfield  {journal}
  {\bibinfo  {journal} {Phys.\ Rev.\ Lett.}\ }\textbf {\bibinfo {volume}
  {91}},\ \bibinfo {pages} {262001} (\bibinfo {year} {2003})},\ \Eprint
  {https://arxiv.org/abs/hep-ex/0309032} {arXiv:hep-ex/0309032} \BibitemShut
  {NoStop}%
\bibitem [{\citenamefont {Lebed}\ \emph {et~al.}(2017)\citenamefont {Lebed},
  \citenamefont {Mitchell},\ and\ \citenamefont {Swanson}}]{Lebed:2016hpi}%
  \BibitemOpen
  \bibfield  {author} {\bibinfo {author} {\bibfnamefont {R.}~\bibnamefont
  {Lebed}}, \bibinfo {author} {\bibfnamefont {R.}~\bibnamefont {Mitchell}},\
  and\ \bibinfo {author} {\bibfnamefont {E.}~\bibnamefont {Swanson}},\ }\href
  {https://doi.org/10.1016/j.ppnp.2016.11.003} {\bibfield  {journal} {\bibinfo
  {journal} {Prog.\ Part.\ Nucl.\ Phys.}\ }\textbf {\bibinfo {volume} {{\bf
  93}}},\ \bibinfo {pages} {143} (\bibinfo {year} {2017})},\ \Eprint
  {https://arxiv.org/abs/1610.04528} {arXiv:1610.04528 [hep-ph]} \BibitemShut
  {NoStop}%
\bibitem [{\citenamefont {Chen}\ \emph {et~al.}(2016)\citenamefont {Chen},
  \citenamefont {Chen}, \citenamefont {Liu},\ and\ \citenamefont
  {Zhu}}]{Chen:2016qju}%
  \BibitemOpen
  \bibfield  {author} {\bibinfo {author} {\bibfnamefont {H.-X.}\ \bibnamefont
  {Chen}}, \bibinfo {author} {\bibfnamefont {W.}~\bibnamefont {Chen}}, \bibinfo
  {author} {\bibfnamefont {X.}~\bibnamefont {Liu}},\ and\ \bibinfo {author}
  {\bibfnamefont {S.-L.}\ \bibnamefont {Zhu}},\ }\href
  {https://doi.org/10.1016/j.physrep.2016.05.004} {\bibfield  {journal}
  {\bibinfo  {journal} {Phys.\ Rep.}\ }\textbf {\bibinfo {volume} {{\bf
  639}}},\ \bibinfo {pages} {1} (\bibinfo {year} {2016})},\ \Eprint
  {https://arxiv.org/abs/1601.02092} {arXiv:1601.02092 [hep-ph]} \BibitemShut
  {NoStop}%
\bibitem [{\citenamefont {Hosaka}\ \emph {et~al.}(2016)\citenamefont {Hosaka},
  \citenamefont {Iijima}, \citenamefont {Miyabayashi}, \citenamefont {Sakai},\
  and\ \citenamefont {Yasui}}]{Hosaka:2016pey}%
  \BibitemOpen
  \bibfield  {author} {\bibinfo {author} {\bibfnamefont {A.}~\bibnamefont
  {Hosaka}}, \bibinfo {author} {\bibfnamefont {T.}~\bibnamefont {Iijima}},
  \bibinfo {author} {\bibfnamefont {K.}~\bibnamefont {Miyabayashi}}, \bibinfo
  {author} {\bibfnamefont {Y.}~\bibnamefont {Sakai}},\ and\ \bibinfo {author}
  {\bibfnamefont {S.}~\bibnamefont {Yasui}},\ }\href
  {https://doi.org/10.1093/ptep/ptw045} {\bibfield  {journal} {\bibinfo
  {journal} {Prog.\ Theor.\ Exp.\ Phys.}\ }\textbf {\bibinfo {volume} {{\bf
  2016}}},\ \bibinfo {pages} {062C01} (\bibinfo {year} {2016})},\ \Eprint
  {https://arxiv.org/abs/1603.09229} {arXiv:1603.09229 [hep-ph]} \BibitemShut
  {NoStop}%
\bibitem [{\citenamefont {Esposito}\ \emph {et~al.}(2017)\citenamefont
  {Esposito}, \citenamefont {Pilloni},\ and\ \citenamefont
  {Polosa}}]{Esposito:2016noz}%
  \BibitemOpen
  \bibfield  {author} {\bibinfo {author} {\bibfnamefont {A.}~\bibnamefont
  {Esposito}}, \bibinfo {author} {\bibfnamefont {A.}~\bibnamefont {Pilloni}},\
  and\ \bibinfo {author} {\bibfnamefont {A.}~\bibnamefont {Polosa}},\ }\href
  {https://doi.org/https://doi.org/10.1016/j.physrep.2016.11.002} {\bibfield
  {journal} {\bibinfo  {journal} {Phys.\ Rep.}\ }\textbf {\bibinfo {volume}
  {\textbf{668}}},\ \bibinfo {pages} {1} (\bibinfo {year} {2017})},\ \Eprint
  {https://arxiv.org/abs/1611.07920} {arXiv:1611.07920 [hep-ph]} \BibitemShut
  {NoStop}%
\bibitem [{\citenamefont {Guo}\ \emph {et~al.}(2018)\citenamefont {Guo},
  \citenamefont {Hanhart}, \citenamefont {Mei{\ss}ner}, \citenamefont {Wang},
  \citenamefont {Zhao},\ and\ \citenamefont {Zou}}]{Guo:2017jvc}%
  \BibitemOpen
  \bibfield  {author} {\bibinfo {author} {\bibfnamefont {F.-K.}\ \bibnamefont
  {Guo}}, \bibinfo {author} {\bibfnamefont {C.}~\bibnamefont {Hanhart}},
  \bibinfo {author} {\bibfnamefont {U.-G.}\ \bibnamefont {Mei{\ss}ner}},
  \bibinfo {author} {\bibfnamefont {Q.}~\bibnamefont {Wang}}, \bibinfo {author}
  {\bibfnamefont {Q.}~\bibnamefont {Zhao}},\ and\ \bibinfo {author}
  {\bibfnamefont {B.-S.}\ \bibnamefont {Zou}},\ }\href
  {https://doi.org/10.1103/RevModPhys.90.015004} {\bibfield  {journal}
  {\bibinfo  {journal} {Rev.\ Mod.\ Phys.}\ }\textbf {\bibinfo {volume} {{\bf
  90}}},\ \bibinfo {pages} {015004} (\bibinfo {year} {2018})},\ \Eprint
  {https://arxiv.org/abs/1705.00141} {arXiv:1705.00141 [hep-ph]} \BibitemShut
  {NoStop}%
\bibitem [{\citenamefont {Ali}\ \emph {et~al.}(2017)\citenamefont {Ali},
  \citenamefont {Lange},\ and\ \citenamefont {Stone}}]{Ali:2017jda}%
  \BibitemOpen
  \bibfield  {author} {\bibinfo {author} {\bibfnamefont {A.}~\bibnamefont
  {Ali}}, \bibinfo {author} {\bibfnamefont {J.}~\bibnamefont {Lange}},\ and\
  \bibinfo {author} {\bibfnamefont {S.}~\bibnamefont {Stone}},\ }\href
  {https://doi.org/10.1016/j.ppnp.2017.08.003} {\bibfield  {journal} {\bibinfo
  {journal} {Prog.\ Part.\ Nucl.\ Phys.}\ }\textbf {\bibinfo {volume} {{\bf
  97}}},\ \bibinfo {pages} {123} (\bibinfo {year} {2017})},\ \Eprint
  {https://arxiv.org/abs/1706.00610} {arXiv:1706.00610 [hep-ph]} \BibitemShut
  {NoStop}%
\bibitem [{\citenamefont {Olsen}\ \emph {et~al.}(2018)\citenamefont {Olsen},
  \citenamefont {Skwarnicki},\ and\ \citenamefont {Zieminska}}]{Olsen:2017bmm}%
  \BibitemOpen
  \bibfield  {author} {\bibinfo {author} {\bibfnamefont {S.}~\bibnamefont
  {Olsen}}, \bibinfo {author} {\bibfnamefont {T.}~\bibnamefont {Skwarnicki}},\
  and\ \bibinfo {author} {\bibfnamefont {D.}~\bibnamefont {Zieminska}},\ }\href
  {https://doi.org/10.1103/RevModPhys.90.015003} {\bibfield  {journal}
  {\bibinfo  {journal} {Rev.\ Mod.\ Phys.}\ }\textbf {\bibinfo {volume} {{\bf
  90}}},\ \bibinfo {pages} {015003} (\bibinfo {year} {2018})},\ \Eprint
  {https://arxiv.org/abs/1708.04012} {arXiv:1708.04012 [hep-ph]} \BibitemShut
  {NoStop}%
\bibitem [{\citenamefont {Karliner}\ \emph {et~al.}(2018)\citenamefont
  {Karliner}, \citenamefont {Rosner},\ and\ \citenamefont
  {Skwarnicki}}]{Karliner:2017qhf}%
  \BibitemOpen
  \bibfield  {author} {\bibinfo {author} {\bibfnamefont {M.}~\bibnamefont
  {Karliner}}, \bibinfo {author} {\bibfnamefont {J.}~\bibnamefont {Rosner}},\
  and\ \bibinfo {author} {\bibfnamefont {T.}~\bibnamefont {Skwarnicki}},\
  }\href {https://doi.org/10.1146/annurev-nucl-101917-020902} {\bibfield
  {journal} {\bibinfo  {journal} {Annu.\ Rev.\ Nucl.\ Part.\ Sci.}\ }\textbf
  {\bibinfo {volume} {{\bf 68}}},\ \bibinfo {pages} {17} (\bibinfo {year}
  {2018})},\ \Eprint {https://arxiv.org/abs/1711.10626} {arXiv:1711.10626
  [hep-ph]} \BibitemShut {NoStop}%
\bibitem [{\citenamefont {Yuan}(2018)}]{Yuan:2018inv}%
  \BibitemOpen
  \bibfield  {author} {\bibinfo {author} {\bibfnamefont {C.-Z.}\ \bibnamefont
  {Yuan}},\ }\href {https://doi.org/10.1142/S0217751X18300181} {\bibfield
  {journal} {\bibinfo  {journal} {Int.\ J. Mod.\ Phys.\ A}\ }\textbf {\bibinfo
  {volume} {{\bf 33}}},\ \bibinfo {pages} {1830018} (\bibinfo {year} {2018})},\
  \Eprint {https://arxiv.org/abs/1808.01570} {arXiv:1808.01570 [hep-ex]}
  \BibitemShut {NoStop}%
\bibitem [{\citenamefont {Liu}\ \emph {et~al.}(2019)\citenamefont {Liu},
  \citenamefont {Chen}, \citenamefont {Chen}, \citenamefont {Liu},\ and\
  \citenamefont {Zhu}}]{Liu:2019zoy}%
  \BibitemOpen
  \bibfield  {author} {\bibinfo {author} {\bibfnamefont {Y.-R.}\ \bibnamefont
  {Liu}}, \bibinfo {author} {\bibfnamefont {H.-X.}\ \bibnamefont {Chen}},
  \bibinfo {author} {\bibfnamefont {W.}~\bibnamefont {Chen}}, \bibinfo {author}
  {\bibfnamefont {X.}~\bibnamefont {Liu}},\ and\ \bibinfo {author}
  {\bibfnamefont {S.-L.}\ \bibnamefont {Zhu}},\ }\href
  {https://doi.org/10.1016/j.ppnp.2019.04.003} {\bibfield  {journal} {\bibinfo
  {journal} {Prog.\ Part.\ Nucl.\ Phys.}\ }\textbf {\bibinfo {volume} {{\bf
  107}}},\ \bibinfo {pages} {237} (\bibinfo {year} {2019})},\ \Eprint
  {https://arxiv.org/abs/1903.11976} {arXiv:1903.11976 [hep-ph]} \BibitemShut
  {NoStop}%
\bibitem [{\citenamefont {Brambilla}\ \emph {et~al.}(2020)\citenamefont
  {Brambilla}, \citenamefont {Eidelman}, \citenamefont {Hanhart}, \citenamefont
  {Nefediev}, \citenamefont {Shen}, \citenamefont {Thomas}, \citenamefont
  {Vairo},\ and\ \citenamefont {Yuan}}]{Brambilla:2019esw}%
  \BibitemOpen
  \bibfield  {author} {\bibinfo {author} {\bibfnamefont {N.}~\bibnamefont
  {Brambilla}}, \bibinfo {author} {\bibfnamefont {S.}~\bibnamefont {Eidelman}},
  \bibinfo {author} {\bibfnamefont {C.}~\bibnamefont {Hanhart}}, \bibinfo
  {author} {\bibfnamefont {A.}~\bibnamefont {Nefediev}}, \bibinfo {author}
  {\bibfnamefont {C.-P.}\ \bibnamefont {Shen}}, \bibinfo {author}
  {\bibfnamefont {C.}~\bibnamefont {Thomas}}, \bibinfo {author} {\bibfnamefont
  {A.}~\bibnamefont {Vairo}},\ and\ \bibinfo {author} {\bibfnamefont {C.-Z.}\
  \bibnamefont {Yuan}},\ }\href {https://doi.org/10.1016/j.physrep.2020.05.001}
  {\bibfield  {journal} {\bibinfo  {journal} {Phys.\ Rept.}\ }\textbf {\bibinfo
  {volume} {{\bf 873}}},\ \bibinfo {pages} {1} (\bibinfo {year} {2020})},\
  \Eprint {https://arxiv.org/abs/1907.07583} {arXiv:1907.07583 [hep-ex]}
  \BibitemShut {NoStop}%
\bibitem [{\citenamefont {Chen}\ \emph {et~al.}(2022)\citenamefont {Chen},
  \citenamefont {Chen}, \citenamefont {Liu}, \citenamefont {Liu},\ and\
  \citenamefont {Zhu}}]{Chen:2022asf}%
  \BibitemOpen
  \bibfield  {author} {\bibinfo {author} {\bibfnamefont {H.-X.}\ \bibnamefont
  {Chen}}, \bibinfo {author} {\bibfnamefont {W.}~\bibnamefont {Chen}}, \bibinfo
  {author} {\bibfnamefont {X.}~\bibnamefont {Liu}}, \bibinfo {author}
  {\bibfnamefont {Y.-R.}\ \bibnamefont {Liu}},\ and\ \bibinfo {author}
  {\bibfnamefont {S.-L.}\ \bibnamefont {Zhu}},\ }\href@noop {} {\  (\bibinfo
  {year} {2022})},\ \Eprint {https://arxiv.org/abs/2204.02649}
  {arXiv:2204.02649 [hep-ph]} \BibitemShut {NoStop}%
\bibitem [{\citenamefont {Zyla}\ \emph {et~al.}(2020)\citenamefont {Zyla} \emph
  {et~al.}}]{Zyla:2020zbs}%
  \BibitemOpen
  \bibfield  {author} {\bibinfo {author} {\bibfnamefont {P.}~\bibnamefont
  {Zyla}} \emph {et~al.} (\bibinfo {collaboration} {Particle Data Group}),\
  }\href {https://doi.org/10.1093/ptep/ptaa104} {\bibfield  {journal} {\bibinfo
   {journal} {PTEP}\ }\textbf {\bibinfo {volume} {{\bf 2020}}},\ \bibinfo
  {pages} {083C01} (\bibinfo {year} {2020})}\BibitemShut {NoStop}%
\bibitem [{\citenamefont {Braaten}\ and\ \citenamefont
  {Kusunoki}(2004)}]{Braaten:2003he}%
  \BibitemOpen
  \bibfield  {author} {\bibinfo {author} {\bibfnamefont {E.}~\bibnamefont
  {Braaten}}\ and\ \bibinfo {author} {\bibfnamefont {M.}~\bibnamefont
  {Kusunoki}},\ }\href {https://doi.org/10.1103/PhysRevD.69.074005} {\bibfield
  {journal} {\bibinfo  {journal} {Phys.\ Rev.\ D}\ }\textbf {\bibinfo {volume}
  {{\bf 69}}},\ \bibinfo {pages} {074005} (\bibinfo {year} {2004})},\ \Eprint
  {https://arxiv.org/abs/hep-ph/0311147} {arXiv:hep-ph/0311147} \BibitemShut
  {NoStop}%
\bibitem [{\citenamefont {Tiesinga}\ \emph {et~al.}(2021)\citenamefont
  {Tiesinga}, \citenamefont {Mohr}, \citenamefont {Newell},\ and\ \citenamefont
  {Taylor}}]{Tiesinga:2021myr}%
  \BibitemOpen
  \bibfield  {author} {\bibinfo {author} {\bibfnamefont {E.}~\bibnamefont
  {Tiesinga}}, \bibinfo {author} {\bibfnamefont {P.}~\bibnamefont {Mohr}},
  \bibinfo {author} {\bibfnamefont {D.}~\bibnamefont {Newell}},\ and\ \bibinfo
  {author} {\bibfnamefont {B.}~\bibnamefont {Taylor}},\ }\href
  {https://doi.org/10.1103/RevModPhys.93.025010} {\bibfield  {journal}
  {\bibinfo  {journal} {Rev.\ Mod.\ Phys.}\ }\textbf {\bibinfo {volume} {{\bf
  93}}},\ \bibinfo {pages} {025010} (\bibinfo {year} {2021})}\BibitemShut
  {NoStop}%
\bibitem [{\citenamefont {Barnes}\ \emph {et~al.}(2005)\citenamefont {Barnes},
  \citenamefont {Godfrey},\ and\ \citenamefont {Swanson}}]{Barnes:2005pb}%
  \BibitemOpen
  \bibfield  {author} {\bibinfo {author} {\bibfnamefont {T.}~\bibnamefont
  {Barnes}}, \bibinfo {author} {\bibfnamefont {S.}~\bibnamefont {Godfrey}},\
  and\ \bibinfo {author} {\bibfnamefont {E.}~\bibnamefont {Swanson}},\ }\href
  {https://doi.org/10.1103/PhysRevD.72.054026} {\bibfield  {journal} {\bibinfo
  {journal} {Phys.\ Rev.\ D}\ }\textbf {\bibinfo {volume} {{\bf 72}}},\
  \bibinfo {pages} {054026} (\bibinfo {year} {2005})},\ \Eprint
  {https://arxiv.org/abs/hep-ph/0505002} {arXiv:hep-ph/0505002 [hep-ph]}
  \BibitemShut {NoStop}%
\bibitem [{\citenamefont {Suzuki}(2005)}]{Suzuki:2005ha}%
  \BibitemOpen
  \bibfield  {author} {\bibinfo {author} {\bibfnamefont {M.}~\bibnamefont
  {Suzuki}},\ }\href {https://doi.org/10.1103/PhysRevD.72.114013} {\bibfield
  {journal} {\bibinfo  {journal} {Phys.\ Rev.\ D}\ }\textbf {\bibinfo {volume}
  {{\bf 72}}},\ \bibinfo {pages} {114013} (\bibinfo {year} {2005})},\ \Eprint
  {https://arxiv.org/abs/hep-ph/0508258} {arXiv:hep-ph/0508258} \BibitemShut
  {NoStop}%
\bibitem [{\citenamefont {Brodsky}\ \emph {et~al.}(2014)\citenamefont
  {Brodsky}, \citenamefont {Hwang},\ and\ \citenamefont
  {Lebed}}]{Brodsky:2014xia}%
  \BibitemOpen
  \bibfield  {author} {\bibinfo {author} {\bibfnamefont {S.}~\bibnamefont
  {Brodsky}}, \bibinfo {author} {\bibfnamefont {D.}~\bibnamefont {Hwang}},\
  and\ \bibinfo {author} {\bibfnamefont {R.}~\bibnamefont {Lebed}},\ }\href
  {https://doi.org/10.1103/PhysRevLett.113.112001} {\bibfield  {journal}
  {\bibinfo  {journal} {Phys.\ Rev.\ Lett.}\ }\textbf {\bibinfo {volume} {{\bf
  113}}},\ \bibinfo {pages} {112001} (\bibinfo {year} {2014})},\ \Eprint
  {https://arxiv.org/abs/1406.7281} {arXiv:1406.7281 [hep-ph]} \BibitemShut
  {NoStop}%
\bibitem [{\citenamefont {Lebed}(2017)}]{Lebed:2017min}%
  \BibitemOpen
  \bibfield  {author} {\bibinfo {author} {\bibfnamefont {R.}~\bibnamefont
  {Lebed}},\ }\href {https://doi.org/10.1103/PhysRevD.96.116003} {\bibfield
  {journal} {\bibinfo  {journal} {Phys.\ Rev.\ D}\ }\textbf {\bibinfo {volume}
  {{\bf 96}}},\ \bibinfo {pages} {116003} (\bibinfo {year} {2017})},\ \Eprint
  {https://arxiv.org/abs/1709.06097} {arXiv:1709.06097 [hep-ph]} \BibitemShut
  {NoStop}%
\bibitem [{\citenamefont {Giron}\ \emph {et~al.}(2019)\citenamefont {Giron},
  \citenamefont {Lebed},\ and\ \citenamefont {Peterson}}]{Giron:2019bcs}%
  \BibitemOpen
  \bibfield  {author} {\bibinfo {author} {\bibfnamefont {J.}~\bibnamefont
  {Giron}}, \bibinfo {author} {\bibfnamefont {R.}~\bibnamefont {Lebed}},\ and\
  \bibinfo {author} {\bibfnamefont {C.}~\bibnamefont {Peterson}},\ }\href
  {https://doi.org/10.1007/JHEP05(2019)061} {\bibfield  {journal} {\bibinfo
  {journal} {J. HIgh Energy Phys.}\ }\textbf {\bibinfo {volume} {{\bf 05}}},\
  \bibinfo {pages} {061} (\bibinfo {year} {2019})},\ \Eprint
  {https://arxiv.org/abs/1903.04551} {arXiv:1903.04551 [hep-ph]} \BibitemShut
  {NoStop}%
\bibitem [{\citenamefont {Giron}\ \emph {et~al.}(2020)\citenamefont {Giron},
  \citenamefont {Lebed},\ and\ \citenamefont {Peterson}}]{Giron:2019cfc}%
  \BibitemOpen
  \bibfield  {author} {\bibinfo {author} {\bibfnamefont {J.}~\bibnamefont
  {Giron}}, \bibinfo {author} {\bibfnamefont {R.}~\bibnamefont {Lebed}},\ and\
  \bibinfo {author} {\bibfnamefont {C.}~\bibnamefont {Peterson}},\ }\href
  {https://doi.org/10.1007/JHEP01(2020)124} {\bibfield  {journal} {\bibinfo
  {journal} {J. High Energy Phys.}\ }\textbf {\bibinfo {volume} {{\bf 01}}},\
  \bibinfo {pages} {124}},\ \Eprint {https://arxiv.org/abs/1907.08546}
  {arXiv:1907.08546 [hep-ph]} \BibitemShut {NoStop}%
\bibitem [{\citenamefont {Giron}\ and\ \citenamefont
  {Lebed}(2020{\natexlab{a}})}]{Giron:2020fvd}%
  \BibitemOpen
  \bibfield  {author} {\bibinfo {author} {\bibfnamefont {J.}~\bibnamefont
  {Giron}}\ and\ \bibinfo {author} {\bibfnamefont {R.}~\bibnamefont {Lebed}},\
  }\href {https://doi.org/10.1103/PhysRevD.101.074032} {\bibfield  {journal}
  {\bibinfo  {journal} {Phys.\ Rev.\ D}\ }\textbf {\bibinfo {volume} {{\bf
  101}}},\ \bibinfo {pages} {074032} (\bibinfo {year} {2020}{\natexlab{a}})},\
  \Eprint {https://arxiv.org/abs/2003.02802} {arXiv:2003.02802 [hep-ph]}
  \BibitemShut {NoStop}%
\bibitem [{\citenamefont {Giron}\ and\ \citenamefont
  {Lebed}(2020{\natexlab{b}})}]{Giron:2020qpb}%
  \BibitemOpen
  \bibfield  {author} {\bibinfo {author} {\bibfnamefont {J.}~\bibnamefont
  {Giron}}\ and\ \bibinfo {author} {\bibfnamefont {R.}~\bibnamefont {Lebed}},\
  }\href {https://doi.org/10.1103/PhysRevD.102.014036} {\bibfield  {journal}
  {\bibinfo  {journal} {Phys.\ Rev.\ D}\ }\textbf {\bibinfo {volume} {{\bf
  102}}},\ \bibinfo {pages} {014036} (\bibinfo {year} {2020}{\natexlab{b}})},\
  \Eprint {https://arxiv.org/abs/2005.07100} {arXiv:2005.07100 [hep-ph]}
  \BibitemShut {NoStop}%
\bibitem [{\citenamefont {Giron}\ \emph {et~al.}(2021)\citenamefont {Giron},
  \citenamefont {Lebed},\ and\ \citenamefont {Martinez}}]{Giron:2021sla}%
  \BibitemOpen
  \bibfield  {author} {\bibinfo {author} {\bibfnamefont {J.}~\bibnamefont
  {Giron}}, \bibinfo {author} {\bibfnamefont {R.}~\bibnamefont {Lebed}},\ and\
  \bibinfo {author} {\bibfnamefont {S.}~\bibnamefont {Martinez}},\ }\href
  {https://doi.org/10.1103/PhysRevD.104.054001} {\bibfield  {journal} {\bibinfo
   {journal} {Phys.\ Rev.\ D}\ }\textbf {\bibinfo {volume} {{\bf 104}}},\
  \bibinfo {pages} {054001} (\bibinfo {year} {2021})},\ \Eprint
  {https://arxiv.org/abs/2106.05883} {arXiv:2106.05883 [hep-ph]} \BibitemShut
  {NoStop}%
\bibitem [{\citenamefont {Giron}\ and\ \citenamefont
  {Lebed}(2021)}]{Giron:2021fnl}%
  \BibitemOpen
  \bibfield  {author} {\bibinfo {author} {\bibfnamefont {J.}~\bibnamefont
  {Giron}}\ and\ \bibinfo {author} {\bibfnamefont {R.}~\bibnamefont {Lebed}},\
  }\href {https://doi.org/10.1103/PhysRevD.104.114028} {\bibfield  {journal}
  {\bibinfo  {journal} {Phys.\ Rev.\ D}\ }\textbf {\bibinfo {volume} {{\bf
  104}}},\ \bibinfo {pages} {114028} (\bibinfo {year} {2021})},\ \Eprint
  {https://arxiv.org/abs/2110.05557} {arXiv:2110.05557 [hep-ph]} \BibitemShut
  {NoStop}%
\bibitem [{\citenamefont {Lebed}(2015)}]{Lebed:2015tna}%
  \BibitemOpen
  \bibfield  {author} {\bibinfo {author} {\bibfnamefont {R.}~\bibnamefont
  {Lebed}},\ }\href {https://doi.org/10.1016/j.physletb.2015.08.032} {\bibfield
   {journal} {\bibinfo  {journal} {Phys.\ Lett.\ B}\ }\textbf {\bibinfo
  {volume} {{\bf 749}}},\ \bibinfo {pages} {454} (\bibinfo {year} {2015})},\
  \Eprint {https://arxiv.org/abs/1507.05867} {arXiv:1507.05867 [hep-ph]}
  \BibitemShut {NoStop}%
\bibitem [{\citenamefont {Born}\ and\ \citenamefont
  {Oppenheimer}(1927)}]{Born:1927boa}%
  \BibitemOpen
  \bibfield  {author} {\bibinfo {author} {\bibfnamefont {M.}~\bibnamefont
  {Born}}\ and\ \bibinfo {author} {\bibfnamefont {R.}~\bibnamefont
  {Oppenheimer}},\ }\href {https://doi.org/10.1002/andp.19273892002} {\bibfield
   {journal} {\bibinfo  {journal} {Ann.\ der Phys.}\ }\textbf {\bibinfo
  {volume} {{\bf 389}}},\ \bibinfo {pages} {457} (\bibinfo {year}
  {1927})}\BibitemShut {NoStop}%
\bibitem [{\citenamefont {Baer}(2006)}]{Baer:2006}%
  \BibitemOpen
  \bibfield  {author} {\bibinfo {author} {\bibfnamefont {M.}~\bibnamefont
  {Baer}},\ }\href@noop {} {\emph {\bibinfo {title} {\it Beyond
  Born-Oppenheimer: Electronic Nonadiabatic Coupling Terms and Conical
  Intersections}}}\ (\bibinfo  {publisher} {Wiley},\ \bibinfo {address} {New
  Jersey},\ \bibinfo {year} {2006})\BibitemShut {NoStop}%
\bibitem [{\citenamefont {Bruschini}\ and\ \citenamefont
  {Gonz\'{a}lez}(2020)}]{Bruschini:2020voj}%
  \BibitemOpen
  \bibfield  {author} {\bibinfo {author} {\bibfnamefont {R.}~\bibnamefont
  {Bruschini}}\ and\ \bibinfo {author} {\bibfnamefont {P.}~\bibnamefont
  {Gonz\'{a}lez}},\ }\href {https://doi.org/10.1103/PhysRevD.102.074002}
  {\bibfield  {journal} {\bibinfo  {journal} {Phys.\ Rev.\ D}\ }\textbf
  {\bibinfo {volume} {{\bf 102}}},\ \bibinfo {pages} {074002} (\bibinfo {year}
  {2020})},\ \Eprint {https://arxiv.org/abs/2007.07693} {arXiv:2007.07693
  [hep-ph]} \BibitemShut {NoStop}%
\bibitem [{\citenamefont {Bruschini}\ and\ \citenamefont
  {Gonz\'{a}lez}(2021{\natexlab{a}})}]{Bruschini:2021cty}%
  \BibitemOpen
  \bibfield  {author} {\bibinfo {author} {\bibfnamefont {R.}~\bibnamefont
  {Bruschini}}\ and\ \bibinfo {author} {\bibfnamefont {P.}~\bibnamefont
  {Gonz\'{a}lez}},\ }\href {https://doi.org/10.1103/PhysRevD.103.074009}
  {\bibfield  {journal} {\bibinfo  {journal} {Phys.\ Rev.\ D}\ }\textbf
  {\bibinfo {volume} {{\bf 103}}},\ \bibinfo {pages} {074009} (\bibinfo {year}
  {2021}{\natexlab{a}})},\ \Eprint {https://arxiv.org/abs/2101.04636}
  {arXiv:2101.04636 [hep-ph]} \BibitemShut {NoStop}%
\bibitem [{\citenamefont {Giron}\ and\ \citenamefont
  {Lebed}(2020{\natexlab{c}})}]{Giron:2020wpx}%
  \BibitemOpen
  \bibfield  {author} {\bibinfo {author} {\bibfnamefont {J.}~\bibnamefont
  {Giron}}\ and\ \bibinfo {author} {\bibfnamefont {R.}~\bibnamefont {Lebed}},\
  }\href {https://doi.org/10.1103/PhysRevD.102.074003} {\bibfield  {journal}
  {\bibinfo  {journal} {Phys.\ Rev.\ D}\ }\textbf {\bibinfo {volume} {{\bf
  102}}},\ \bibinfo {pages} {074003} (\bibinfo {year} {2020}{\natexlab{c}})},\
  \Eprint {https://arxiv.org/abs/2008.01631} {arXiv:2008.01631 [hep-ph]}
  \BibitemShut {NoStop}%
\bibitem [{\citenamefont {Bulava}\ \emph {et~al.}(2019)\citenamefont {Bulava},
  \citenamefont {H{\"o}rz}, \citenamefont {Knechtli}, \citenamefont {Koch},
  \citenamefont {Moir}, \citenamefont {Morningstar},\ and\ \citenamefont
  {Peardon}}]{Bulava:2019iut}%
  \BibitemOpen
  \bibfield  {author} {\bibinfo {author} {\bibfnamefont {J.}~\bibnamefont
  {Bulava}}, \bibinfo {author} {\bibfnamefont {B.}~\bibnamefont {H{\"o}rz}},
  \bibinfo {author} {\bibfnamefont {F.}~\bibnamefont {Knechtli}}, \bibinfo
  {author} {\bibfnamefont {V.}~\bibnamefont {Koch}}, \bibinfo {author}
  {\bibfnamefont {G.}~\bibnamefont {Moir}}, \bibinfo {author} {\bibfnamefont
  {C.}~\bibnamefont {Morningstar}},\ and\ \bibinfo {author} {\bibfnamefont
  {M.}~\bibnamefont {Peardon}},\ }\href
  {https://doi.org/10.1016/j.physletb.2019.05.018} {\bibfield  {journal}
  {\bibinfo  {journal} {Phys.\ Lett.\ B}\ }\textbf {\bibinfo {volume} {{\bf
  793}}},\ \bibinfo {pages} {493} (\bibinfo {year} {2019})},\ \Eprint
  {https://arxiv.org/abs/1902.04006} {arXiv:1902.04006 [hep-lat]} \BibitemShut
  {NoStop}%
\bibitem [{\citenamefont {Berwein}\ \emph {et~al.}(2015)\citenamefont
  {Berwein}, \citenamefont {Brambilla}, \citenamefont
  {Tarr{\'u}s~Castell{\`a}},\ and\ \citenamefont {Vairo}}]{Berwein:2015vca}%
  \BibitemOpen
  \bibfield  {author} {\bibinfo {author} {\bibfnamefont {M.}~\bibnamefont
  {Berwein}}, \bibinfo {author} {\bibfnamefont {N.}~\bibnamefont {Brambilla}},
  \bibinfo {author} {\bibfnamefont {J.}~\bibnamefont
  {Tarr{\'u}s~Castell{\`a}}},\ and\ \bibinfo {author} {\bibfnamefont
  {A.}~\bibnamefont {Vairo}},\ }\href
  {https://doi.org/10.1103/PhysRevD.92.114019} {\bibfield  {journal} {\bibinfo
  {journal} {Phys.\ Rev.\ D}\ }\textbf {\bibinfo {volume} {{\bf 92}}},\
  \bibinfo {pages} {114019} (\bibinfo {year} {2015})},\ \Eprint
  {https://arxiv.org/abs/1510.04299} {arXiv:1510.04299 [hep-ph]} \BibitemShut
  {NoStop}%
\bibitem [{\citenamefont {Prkacin}\ \emph {et~al.}(2006)\citenamefont
  {Prkacin}, \citenamefont {Bali}, \citenamefont {Dussel}, \citenamefont
  {Lippert}, \citenamefont {Neff},\ and\ \citenamefont
  {Schilling}}]{Bali:2005}%
  \BibitemOpen
  \bibfield  {author} {\bibinfo {author} {\bibfnamefont {Z.}~\bibnamefont
  {Prkacin}}, \bibinfo {author} {\bibfnamefont {G.}~\bibnamefont {Bali}},
  \bibinfo {author} {\bibfnamefont {T.}~\bibnamefont {Dussel}}, \bibinfo
  {author} {\bibfnamefont {T.}~\bibnamefont {Lippert}}, \bibinfo {author}
  {\bibfnamefont {H.}~\bibnamefont {Neff}},\ and\ \bibinfo {author}
  {\bibfnamefont {K.}~\bibnamefont {Schilling}},\ }\href
  {https://doi.org/10.22323/1.020.0308} {\bibfield  {journal} {\bibinfo
  {journal} {PoS}\ }\textbf {\bibinfo {volume} {{\bf LAT2005}}},\ \bibinfo
  {pages} {308} (\bibinfo {year} {2006})},\ \Eprint
  {https://arxiv.org/abs/hep-lat/0510051} {arXiv:hep-lat/0510051} \BibitemShut
  {NoStop}%
\bibitem [{\citenamefont {Yamaguchi}\ \emph {et~al.}(2020)\citenamefont
  {Yamaguchi}, \citenamefont {Hosaka}, \citenamefont {Takeuchi},\ and\
  \citenamefont {Takizawa}}]{Yamaguchi:2019vea}%
  \BibitemOpen
  \bibfield  {author} {\bibinfo {author} {\bibfnamefont {Y.}~\bibnamefont
  {Yamaguchi}}, \bibinfo {author} {\bibfnamefont {A.}~\bibnamefont {Hosaka}},
  \bibinfo {author} {\bibfnamefont {S.}~\bibnamefont {Takeuchi}},\ and\
  \bibinfo {author} {\bibfnamefont {M.}~\bibnamefont {Takizawa}},\ }\href
  {https://doi.org/10.1088/1361-6471/ab72b0} {\bibfield  {journal} {\bibinfo
  {journal} {J. Phys.\ G}\ }\textbf {\bibinfo {volume} {{\bf 47}}},\ \bibinfo
  {pages} {053001} (\bibinfo {year} {2020})},\ \Eprint
  {https://arxiv.org/abs/1908.08790} {arXiv:1908.08790 [hep-ph]} \BibitemShut
  {NoStop}%
\bibitem [{\citenamefont {Gens}\ \emph {et~al.}(2021)\citenamefont {Gens},
  \citenamefont {Giron},\ and\ \citenamefont {Lebed}}]{Gens:2021wyf}%
  \BibitemOpen
  \bibfield  {author} {\bibinfo {author} {\bibfnamefont {J.}~\bibnamefont
  {Gens}}, \bibinfo {author} {\bibfnamefont {J.}~\bibnamefont {Giron}},\ and\
  \bibinfo {author} {\bibfnamefont {R.}~\bibnamefont {Lebed}},\ }\href
  {https://doi.org/10.1103/PhysRevD.103.094024} {\bibfield  {journal} {\bibinfo
   {journal} {Phys.\ Rev.\ D}\ }\textbf {\bibinfo {volume} {{\bf 103}}},\
  \bibinfo {pages} {094024} (\bibinfo {year} {2021})},\ \Eprint
  {https://arxiv.org/abs/2102.12591} {arXiv:2102.12591 [hep-ph]} \BibitemShut
  {NoStop}%
\bibitem [{\citenamefont {Swanson}(2004)}]{Swanson:2004pp}%
  \BibitemOpen
  \bibfield  {author} {\bibinfo {author} {\bibfnamefont {E.}~\bibnamefont
  {Swanson}},\ }\href {https://doi.org/10.1016/j.physletb.2004.07.059}
  {\bibfield  {journal} {\bibinfo  {journal} {Phys.\ Lett.\ B}\ }\textbf
  {\bibinfo {volume} {{\bf 598}}},\ \bibinfo {pages} {197} (\bibinfo {year}
  {2004})},\ \Eprint {https://arxiv.org/abs/hep-ph/0406080}
  {arXiv:hep-ph/0406080} \BibitemShut {NoStop}%
\bibitem [{\citenamefont {Guo}\ \emph {et~al.}(2015)\citenamefont {Guo},
  \citenamefont {Hanhart}, \citenamefont {Kalashnikova}, \citenamefont
  {Mei\ss{}ner},\ and\ \citenamefont {Nefediev}}]{Guo:2014taa}%
  \BibitemOpen
  \bibfield  {author} {\bibinfo {author} {\bibfnamefont {F.-K.}\ \bibnamefont
  {Guo}}, \bibinfo {author} {\bibfnamefont {C.}~\bibnamefont {Hanhart}},
  \bibinfo {author} {\bibfnamefont {Y.~S.}\ \bibnamefont {Kalashnikova}},
  \bibinfo {author} {\bibfnamefont {U.-G.}\ \bibnamefont {Mei\ss{}ner}},\ and\
  \bibinfo {author} {\bibfnamefont {A.}~\bibnamefont {Nefediev}},\ }\href
  {https://doi.org/10.1016/j.physletb.2015.02.013} {\bibfield  {journal}
  {\bibinfo  {journal} {Phys.\ Lett.\ B}\ }\textbf {\bibinfo {volume} {{\bf
  742}}},\ \bibinfo {pages} {394} (\bibinfo {year} {2015})},\ \Eprint
  {https://arxiv.org/abs/1410.6712} {arXiv:1410.6712 [hep-ph]} \BibitemShut
  {NoStop}%
\bibitem [{\citenamefont {Cincioglu}\ and\ \citenamefont
  {Ozpineci}(2019)}]{Cincioglu:2019gzd}%
  \BibitemOpen
  \bibfield  {author} {\bibinfo {author} {\bibfnamefont {E.}~\bibnamefont
  {Cincioglu}}\ and\ \bibinfo {author} {\bibfnamefont {A.}~\bibnamefont
  {Ozpineci}},\ }\href {https://doi.org/10.1016/j.physletb.2019.134856}
  {\bibfield  {journal} {\bibinfo  {journal} {Phys.\ Lett.\ B}\ }\textbf
  {\bibinfo {volume} {{\bf 797}}},\ \bibinfo {pages} {134856} (\bibinfo {year}
  {2019})},\ \Eprint {https://arxiv.org/abs/1901.03138} {arXiv:1901.03138
  [hep-ph]} \BibitemShut {NoStop}%
\bibitem [{\citenamefont {Liu}\ \emph {et~al.}(2008)\citenamefont {Liu},
  \citenamefont {Liu}, \citenamefont {Deng},\ and\ \citenamefont
  {Zhu}}]{Liu:2008fh}%
  \BibitemOpen
  \bibfield  {author} {\bibinfo {author} {\bibfnamefont {Y.-R.}\ \bibnamefont
  {Liu}}, \bibinfo {author} {\bibfnamefont {X.}~\bibnamefont {Liu}}, \bibinfo
  {author} {\bibfnamefont {W.-Z.}\ \bibnamefont {Deng}},\ and\ \bibinfo
  {author} {\bibfnamefont {S.-L.}\ \bibnamefont {Zhu}},\ }\href
  {https://doi.org/10.1140/epjc/s10052-008-0640-4} {\bibfield  {journal}
  {\bibinfo  {journal} {Eur.\ Phys.\ J. C}\ }\textbf {\bibinfo {volume} {{\bf
  56}}},\ \bibinfo {pages} {63} (\bibinfo {year} {2008})},\ \Eprint
  {https://arxiv.org/abs/0801.3540} {arXiv:0801.3540 [hep-ph]} \BibitemShut
  {NoStop}%
\bibitem [{\citenamefont {Bruschini}\ and\ \citenamefont
  {Gonz\'{a}lez}(2021{\natexlab{b}})}]{Bruschini:2021ckr}%
  \BibitemOpen
  \bibfield  {author} {\bibinfo {author} {\bibfnamefont {R.}~\bibnamefont
  {Bruschini}}\ and\ \bibinfo {author} {\bibfnamefont {P.}~\bibnamefont
  {Gonz\'{a}lez}},\ }\href {https://doi.org/10.1103/PhysRevD.104.074025}
  {\bibfield  {journal} {\bibinfo  {journal} {Phys.\ Rev.\ D}\ }\textbf
  {\bibinfo {volume} {{\bf 104}}},\ \bibinfo {pages} {074025} (\bibinfo {year}
  {2021}{\natexlab{b}})},\ \Eprint {https://arxiv.org/abs/2107.05459}
  {arXiv:2107.05459 [hep-ph]} \BibitemShut {NoStop}%
\end{thebibliography}%
\end{document}